\documentclass[aps,prx,superscriptaddress,twocolumn,10pt]{revtex4-2}

\usepackage[titletoc,toc,title,page]{appendix}

\usepackage{csquotes}
\usepackage[italian,english]{babel}

\usepackage{microtype} 
\usepackage[margin=.6in]{geometry}

\usepackage[sc]{mathpazo} 

\usepackage{parskip}
\usepackage{bm} 

\usepackage{dsfont} 
\usepackage{amsmath,amssymb,amsthm,thmtools}
\usepackage{mathtools}
\usepackage{cases}
\usepackage{calc}
\usepackage{mathrsfs} 
\usepackage[normalem]{ulem} 
\usepackage{nameref}
\usepackage[colorlinks=true]{hyperref}
\usepackage[nameinlink]{cleveref}
\crefname{appsec}{Appendix}{Appendices}
\crefname{box}{Box}{Box}
\hypersetup{
  colorlinks   = true, 
  urlcolor     = green!80!black, 
  linkcolor    = blue, 
  citecolor    = red!80!black, 
  breaklinks=true   
}

\usepackage{physics} 

\usepackage{float} 

\usepackage{graphicx}

\usepackage[usenames,dvipsnames,table]{xcolor}
\usepackage{easyReview}

\usepackage{tikz}
\usetikzlibrary{calc,shapes.geometric}

\usepackage{multirow,tabularx,booktabs}
\usepackage{makecell}

\newcommand{\EE}{\mathbb{E}}
\newcommand{\PP}{\mathbb{P}}

\newcommand{\calC}{\mathcal{C}}

\newcommand{\calE}{\mathcal{E}}
\newcommand{\calF}{\mathcal{F}}
\newcommand{\calM}{\mathcal{M}}
\newcommand{\calU}{\mathcal{U}}
\newcommand{\calO}{\mathcal{O}}

\newcommand{\calH}{\mathcal{H}}

\newcommand{\bs}[1]{\boldsymbol{#1}}
\newcommand{\on}[1]{\operatorname{#1}}
\newcommand{\parTitle}[1]{\noindent\emph{#1} --- }

\definecolor{azure}{rgb}{0.0, 0.3, 1.0}

\makeatletter
\newcommand{\labeltarget}[1]{\Hy@raisedlink{\hypertarget{#1}{}}}
\makeatother
\begin{document}
\title{Shadow tomography on general measurement frames}
\author{L. Innocenti}
\let\comma,
\affiliation{Universit\`a degli Studi di Palermo\comma{} Dipartimento di Fisica e Chimica - Emilio Segr\`e\comma{} via Archirafi 36\comma{} I-90123 Palermo\comma{} Italy}
\author{S. Lorenzo}
\affiliation{Universit\`a degli Studi di Palermo\comma{} Dipartimento di Fisica e Chimica - Emilio Segr\`e\comma{} via Archirafi 36\comma{} I-90123 Palermo\comma{} Italy}
\author{I. Palmisano}
\affiliation{Centre for Quantum Materials and Technologies\comma{} School of Mathematics and Physics\comma{} Queen's University Belfast\comma{} BT7 1NN\comma{} United Kingdom}
\author{F. Albarelli}
\affiliation{Quantum Technology Lab\comma{} Dipartimento di Fisica Aldo Pontremoli\comma{} Universit\`a degli Studi di Milano\comma{} I-20133 Milano\comma{} Italy}
\affiliation{Istituto Nazionale di Fisica Nucleare\comma{} Sezione di Milano\comma{} via Celoria 16\comma{}  20133 Milan\comma{} Italy}
\author{A. Ferraro}
\affiliation{Centre for Quantum Materials and Technologies\comma{} School of Mathematics and Physics\comma{} Queen's University Belfast\comma{} BT7 1NN\comma{} United Kingdom}
\affiliation{Quantum Technology Lab\comma{} Dipartimento di Fisica Aldo Pontremoli\comma{} Universit\`a degli Studi di Milano\comma{} I-20133 Milano\comma{} Italy} 
\author{M. Paternostro}
\affiliation{Centre for Quantum Materials and Technologies\comma{} School of Mathematics and Physics\comma{} Queen's University Belfast\comma{} BT7 1NN\comma{} United Kingdom}
\affiliation{Universit\`a degli Studi di Palermo\comma{} Dipartimento di Fisica e Chimica - Emilio Segr\`e\comma{} via Archirafi 36\comma{} I-90123 Palermo\comma{} Italy}
\author{G. M. Palma}
\affiliation{Universit\`a degli Studi di Palermo\comma{} Dipartimento di Fisica e Chimica - Emilio Segr\`e\comma{} via Archirafi 36\comma{} I-90123 Palermo\comma{} Italy}
\affiliation{NEST\comma{} Istituto Nanoscienze-CNR\comma{} Piazza S. Silvestro 12\comma{} 56127 Pisa\comma{} Italy}

\begin{abstract}
    We provide a new perspective on shadow tomography by demonstrating its deep connections with the general theory of measurement frames. By showing that the formalism of measurement frames offers a natural framework for shadow tomography --- in which ``classical shadows'' correspond to unbiased estimators derived from a suitable dual frame associated with the given measurement --- we highlight the intrinsic connection between standard state tomography and shadow tomography. Such perspective allows us to examine the interplay between measurements, reconstructed observables, and the estimators used to process measurement outcomes, while paving the way to assess the influence of the input state and the dimension of the underlying space on estimation errors. Our approach generalizes the method described in [H.-Y. Huang {\it et al.}, Nat. Phys. {\bf 16}, 1050 (2020)], whose results are recovered in the special case of covariant measurement frames. As an application, we demonstrate that a sought-after target of shadow tomography can be achieved for the entire class of tight rank-1 measurement frames --- namely, that it is possible to accurately estimate a finite set of generic rank-1 bounded observables while avoiding the growth of the number of the required samples with the state dimension.
\end{abstract}
\maketitle

\section{Introduction}

The reliable reconstruction of the information encoded in a quantum register is one of the stepping stones of any quantum information processing device. In this respect, quantum state tomography (QST), that is, the task of estimating quantum states from a measured dataset, is the gold standard for verification and benchmarking of quantum devices~\cite{paris2004lectures,teo2015introduction,dariano2003quantum}. QST has been performed in countless experiments by measuring a complete set of observables whose expectation values determine the quantum state.

As the typical representation of density matrices implies a number of coefficients exponential in the number of constituent subsystems, the standard formulation of tomography~\cite{james2001measurement} of a generic state requires an exponential time in the system size. Alternative methods
based on efficient representations of multiparty quantum states -- such as matrix product states~\cite{MPS} -- have led to improved schemes for state tomography. Such an advantage, however, is achieved only for those states that are efficiently represented in the ansatz that is chosen. 
On the other hand, performing QST of $d$-dimensional quantum states, within error $\epsilon$ (in trace distance), requires a number of copies of the unknown state that scales polynomially with $d$~\cite{james2001measurement,teo2015introduction}. In this context, tight lower bounds to single-copy non-adaptive state reconstruction have been proven
~\cite{lowe2022lower,guctua2020fast,kueng2014low,haah2016sample}.

However, reconstructing specific features of a state, rather than performing full tomographic reconstruction, is achievable with a much smaller amount of resources~\cite{morris2021quantum,gebhart2023learning}.
In particular, the number of measurements required to estimate the expectation value of $M$ observables within error $\epsilon$ scales logarithmically with $M$, and does not depend explicitly on the stace dimension --- the associated task is referred to as ``shadow tomography''~\cite{aaronson2018shadow}.
An explicit way to implement shadow tomography via random Clifford circuits was recently proposed~\cite{huang2019predicting,huang2020predicting,elben2022review}.
A review discussing some of the relations between state tomography and shadow tomography is found in Ref.~\cite{kliesch2021theory}.
In particular, a generalization of shadow tomography to general quantum measurements was recently proposed in Ref.~\cite{nguyen2022optimising,acharya2021informationally}.

Here, we further the grounding of shadow tomography for agile property reconstruction by highlighting its deep connection with the approach of state tomography via measurement frames~\cite{scott2006tight,dariano2007optimal,perinotti2007optimal,ferrie2008frame,zhu2014quantum,dariano2007optimal,perez2022mutually,fuchs2017sic}.
Our formalism reduces to the standard approach of~\cite{huang2020predicting} in special cases, and is compatible with its generalizations presented in~\cite{acharya2021informationally,nguyen2022optimising}.
We demonstrate that this general formalism provides a simple framework to understand the 
relationship between measurement, target observable, and estimator used to post-process measurement outcomes, as well as how the input state and the dimension of the underlying space affect the estimation error.
This approach also directly connects with general metrological considerations, showing how classical shadows can be seen as minimum-variance unbiased linear estimators.
This formalism can also potentially be of great use to study the efficiency of state estimation schemes involving generalized measurements and single-setting measurement schemes, which have recently attracted significant attention~\cite{stricker2022experimental,garcia2021learning}.

More specifically, we take the analysis of measurement frames developed for state tomography, and specialize it to analyze estimation errors for shadow tomography tasks.
We discuss how the mean squared erorr (MSE) matrix, a quantity defined to study state tomography whose trace gives estimation error, also reveals a powerful tool to study errors in shadow tomography.
We show how, for any choice of measurement, multiple possible unbiased estimators can be used to post-process the measurement data to recover the target observables, and discuss how to find the unbiased estimator that minimizes the variance with respect to any given input state, as well as the one with minimum averaged variance --- with the average taken with respect to uniformly random input states.
We also demonstrate that the notion of \emph{shadow norm} of an observable, introduced in Ref.~\cite{huang2020predicting}, emerges naturally in this more general formalism.
Furthermore, we examine the behavior of errors for different choices of measurement, given a fixed optimal estimator.
A crucial feature of shadow tomography is the favorable scaling of estimation errors with the dimension of the state.
Focusing on this aspect, we derive explicit bounds for best- and worst-case estimation errors corresponding to different measurement choices, and find a wide class of measurements which allows to estimate properties as efficiently as the protocol used in Ref.~\cite{huang2020predicting}.

\parTitle{Outline} The remainder of this manuscript is organized as follows. In~\cref{sec:shadow_tomography_formalism} we present a reformulation of shadow tomography using the formalism of measurement frames.
In~\cref{sec:estimators_derivation} we introduce the notion of canonical estimator, review standard results for linear tomography in the measurement frames formalism, and highlight the strong analogy between shadow and linear tomography.
In~\cref{sec:bounds_averaged_variance} we derive general bounds for the variance of the introduced estimators, both in the averaged and best- and worst-case settings, and establish general results connecting the symmetry of the measurement with the associated variances.
In~\cref{sec:original_formalism} we show explicitly how the formalism introduced in~\cite{huang2020predicting} can be viewed as a special instance of our approach, specifically when employing covariant measurements and canonical estimators.
Conclusions and forward looks are finally given in~\cref{sec:conclusions}.
Additional in-depth discussions about the derivations and formalism used throughout the paper can be found in the appendices.

\section{Shadow tomography on measurement frames}
\label{sec:shadow_tomography_formalism}

In this section, we demonstrate explicitly how the formalism of measurement frames provides a natural framework for discussing shadow tomography on general quantum measurements.
The approach to shadow tomography~\cite{aaronson2018shadow} introduced in~\cite{huang2019predicting,huang2020predicting} relies on the idea of \textit{classical shadows}, which are functions of the measurement outcomes that can be used to derive good estimates for target observables.
These classical shadows can be understood as a way to construct unbiased estimators for the input state that operate on individual measurement outcomes.
Unbiased estimators for target observables are then easily obtained via these classical shadows.
By not requiring to recover a tomographically complete description of the states, such specialized estimators allow to efficiently estimate desired features of input states.
An explicit protocol to perform shadow tomography with Clifford circuits was recently proposed in~\cite{huang2019predicting,huang2020predicting}, and some generalizations to general measurements were proposed in~\cite{acharya2021informationally,nguyen2022optimising}.
Here, we demonstrate that frame theory~\cite{casazza2013introduction,casazza2015brief}, and in particular the formalism of measurement frames~\cite{scott2006tight,zhu2011quantum,zhu2012phdthesis,zhu2014tomographic}, provide a remarkably simple conceptual framework to think about shadow tomography, and allow to directly view the ``classical shadows'' as the unbiased estimators which constitute the elements of the dual measurement frame.

\parTitle{Notation}
We will restrict our attention to finite-dimensional states and measurements with a finite number of outcomes. This constraint allows a more concise presentation, and can relaxed later without significantly changing the formalism or the results. Following the notation of~\cite{watrous2018theory}, we will denote the real vector space of Hermitian operators acting on a $d$-dimensional complex vector space $\mathbb{C}^d$ as $\on{Herm}(\mathbb{C}^d)$, the set of positive semidefinite operators acting on the same space as $\on{Pos}(\mathbb{C}^d)$, and the subset of density matrices as $\mathrm D(\mathbb{C}^d)\subset\operatorname{Pos}(\mathbb{C}^d)$.
To focus on the linear algebraic properties involved in the calculations, we will use the notation $\langle X,Y\rangle\equiv\tr(X^\dagger Y)$ to denote the Hilbert-Schmidt inner product between operators $X,Y$, and $\| X \|_{2}\equiv \sqrt{\operatorname{tr}(X^2)}$ for the corresponding operator norm.
We will denote a positive operator-valued measure (POVM) with $\ell$ outcomes as $\bs\mu\equiv(\mu_a)_{a=1}^\ell$, where $\mu_a\in\on{Pos}(\mathbb{C}^d)$ and $\sum_a \mu_a=I$. Given a state $\rho\in\mathrm D(\mathbb{C}^d)$, the associated outcome probabilities are thus given by $p_a(\rho)=\langle\mu_a,\rho\rangle$.
Any procedure involving an arbitrary evolution followed by a measurement in some basis can be concisely modeled via one such POVM.

\parTitle{Frame theory}
In linear algebra, a \textit{frame}~\cite{christensen2003introduction,casazza2013introduction,casazza2015brief} for a vector space $V$ is a collection of vectors $v_k\in V$ such that, for all $v\in V$,
$A\|v\|^2 \le \sum_k |\langle v_k,v\rangle|^2\le B\|v\|^2$,
for some $0< A\le B<\infty$.
These can informally be thought of as overcomplete bases: sets of vectors spanning the space, thus providing a linear decomposition for all other vectors.
For finite frames in finite-dimensional spaces, a set $(v_k)_k$ is a frame iff it spans $V$~\cite{casazza2013introduction}.
Given a frame $(v_k)_k$, any $v\in V$ can be linearly decomposed as
\begin{equation}
    v = \sum_k \langle v_k,v\rangle \tilde v_k
    = \sum_k \langle \tilde v_k,v\rangle v_k,
\end{equation}
where $(\tilde v_k)_k$ is another frame, referred to as a \textit{dual frame} of $(v_k)_k$. A frame $(v_k)_k$ admits infinitely many possible dual frames iff it is not linearly independent --- \textit{i.e.} if it is ``overcomplete''.

If we want to estimate a given unknown state $\rho$ from measurement outcomes, a natural class of objects to study are \textit{unbiased estimators}.
These are functions $\hat f:\Sigma\to\on{Herm}(\mathbb{C}^d)$, which map the set of measurement outcomes $\Sigma$ into Hermitian operators which on average reproduce the measured state. That is, more precisely:
\begin{equation}\label{eq:unbiased_estimator_def}
    \mathbb{E}[\hat f|\rho] \equiv
    \sum_a \langle\mu_a,\rho\rangle \hat f(a) = \rho.
\end{equation}
The elements of a POVM $\bs\mu\equiv(\mu_a)_{a\in\Sigma}$ are vectors in $\on{Herm}(\mathbb{C}^d)$, and span linearly the space of Hermitian operators iff they are informationally complete (IC)~\cite{watrous2018theory}.
We can therefore think of $\bs\mu$ as a frame of operators in the real space $\on{Herm}(\mathbb{C}^d)$ equipped with the Hilbert-Schmidt inner product. Such frames of operators are referred to as \textit{measurement frames}~\cite{scott2006tight,zhu2014quantum,dariano2007optimal,perinotti2007optimal,eldar2002optimal,bisio2009}. The task of finding unbiased estimators is thus equivalent to that of finding dual measurement frames for a given IC-POVM $\bs\mu$.
A natural choice of dual frame is the \textit{canonical dual frame} $(\mu_a^\star)_{a\in \Sigma}$ defined via the \textit{frame superoperator} $\calF\in\on{Lin}(\on{Herm}(\mathbb{C}^d))$ as
\begin{equation}\label{eq:frame_operator_def}
    \mu_a^\star \equiv \calF^{-1}(\mu_a),
    \qquad
    \calF(X) \equiv \sum_{a} \langle\mu_a,X\rangle \mu_a.
\end{equation}
This definition of canonical dual frame is a direct application of the standard procedure used in frame theory for generic frames of vectors, where one can define a \textit{frame operator} that, acting on frame elements, gives the corresponding canonical dual frame elements.
Here, the vectors making up the frame are operators themselves. Therefore, in our context, such frame operators are linear operators acting on operators. We will refer to this type of linear transformations as frame \textit{super}operator in order to highlight such technical aspect. Equivalently, $\calF$ and $\calF^{-1}$ can be thought of as quantum maps, which linearly transforms operators into other operators.
The frame superoperator can also be concisely written as
$\calF = \sum_a \mathbb{P}(\mu_a)$,
where $\mathbb{P}(Y)\in\on{Pos}(\on{Herm}(\mathbb{C}^d))$ denotes the outer product of $Y \in \on{Herm}(\mathbb{C}^d)$ with itself, \textit{i.e.} the superoperator acting as $\mathbb{P}(Y): \rho\mapsto\langle Y,\rho\rangle Y$ on any $\rho\in \on{Herm}(\mathbb{C}^d)$.
In vectorized bra-ket notation, this is also often denoted as $\mathbb{P}(Y)\equiv |Y\rangle\!\rangle\!\langle\!\langle Y|$.
Note that $\mathbb{P}(Y)$ is therefore again a quantum map, and its action on an operator $\rho$ would thus read explicitly $\mathbb{P}(Y)(\rho)=\langle Y,\rho\rangle Y\equiv \operatorname{tr}(Y^\dagger \rho)Y$.
There are in general infinitely many dual frames associated with any given $\bs\mu$, each one corresponding to a different unbiased estimator. These estimators are not generally equivalent, and can result in different reconstruction efficiencies. This will be discussed in detail in~\cref{sec:estimators_derivation}.
In particular, while $(\mu_a^\star)_{a\in \Sigma}$ is a standard choice of dual in the context of frame theory, we will show that it is not in fact the optimal choice to estimate properties of input states.

\parTitle{Estimators from measurement frames}
In summary, for any IC-POVM $\bs\mu$ and dual measurement frame $\tilde{\bs\mu}$, we have an unbiased estimator $\hat f(b)\equiv \tilde\mu_b$ for the unknown input state $\rho$, and vice versa, any such unbiased estimator can be obtained from a dual measurement frame of $\bs\mu$.
If the goal is estimating the expectation value of an observable $\calO$, we use the estimator $\hat o(b)\equiv \langle\calO,\hat f(b)\rangle$.
With this formalism, we can understand the main scaling results of shadow tomography as the observation that by carefully choosing the measurement $\bs\mu$ and associated dual measurement frame $\tilde{\bs\mu}$, we obtain favorable scalings to estimate (finite sets of) target observables.
The connection with the standard framing of shadow tomography, is that the \textit{classical shadows} are precisely a particular --- in some sense optimal --- choice of state estimators $\hat f$.
If a finite set of outcomes $\{b_1,...,b_N\}$ is collected, we compute and store the values of the single-outcome estimators $\hat f(b_k)$, and then build from these an estimator for the expectation value --- typically via the sample mean $\frac1N\sum_{k=1}^N \hat f(b_k)$, or median-of-means.
To estimate the expectation value of $\calO$, the average is instead computed on the values $\langle\calO,\hat f(b)\rangle$.

\parTitle{Variance of the estimators}
A standard way to assess the magnitude of the statistical fluctuations in the estimator is to consider its variance.
For state estimators, considering the errors in $L_2$ distance, the variance reads
\begin{equation}
    \on{Var}[\hat f] =
    \EE[\|\hat f - \rho\|_2^2] =
    \sum_b \langle \mu_b,\rho\rangle
    \|\hat f(b) - \rho\|^2_2.
\end{equation}
Similarly, for observable estimators, the variance reads
\begin{equation}\small
    \on{Var}[\hat o] =
    \mathbb{E}[(\hat o - \langle\calO,\rho\rangle)^2] =
    \sum_b \langle\mu_b,\rho\rangle (\hat o(b) - \langle\calO,\rho\rangle)^2.
\end{equation}
These variances depend on input state $\rho$, measurement $\bs\mu$, estimator $\hat f$, and target observable $\calO$.
For the sake of conciseness, the dependence on some or all of these will often not be made explicit, using the shorthand $\on{Var}[\hat o]\equiv\on{Var}[\hat o|\rho,\bs\mu,\hat f,\calO]$.
Knowledge of the variance grants performance guarantees for the additive estimation error, via standard statistical bounds such as Chebyshev's, Hoeffding's, or Bernstein's inequalities, or employing median-of-means estimators. A recent discussion of these statistical bounds and their applications to quantum state estimation is given in~\cite{kliesch2021theory}.
As will be shown in detail in the following sections, for the entire class of so-called ``tight measurement frames'', we can derive the unbiased estimator that minimizes the averaged variance, and show that its averaged variance does not depend explicitly on the state dimension. Furthermore, for any measurement frame that forms a 3-design, we will prove that also the worst-case scenario variance can be similarly upper bounded. This generalizes some of the results reported in~\cite{huang2020predicting} for random measurements.

\parTitle{Non-positivity of state estimators}
It is worth noting that the state estimators $\hat f(b)$ obtained with this scheme are Hermitian matrices, but not necessarily have unit trace or are positive semidefinite. This means that if the goal is to estimate the state itself, the estimated state might not be a valid density matrix. This is precisely what happens in the context of linear state tomography, and is also the defining setting of shadow tomography. This feature of the scheme is particularly not problematic in the shadow tomography setting because the focus is on reconstructing expectation values of observables, rather than the density matrix itself.

\parTitle{Mean vs median-of-means estimators}
The median-of-means estimator, which was used for example in~\cite{huang2020predicting}, was recently found to not provide an advantage over the standard mean estimator in some situations~\cite{helsen2022thrifty,acharya2021informationally}.
More generally, Hoeffding-like bounds provide the same scaling performance guarantees for any sub-Gaussian distribution, and thus in particular for bounded ones~\cite{devroye2016sub}.
All the estimators for finite-dimensional observables we study are bounded by construction: for any IC-POVM $\bs\mu$, estimator $\bs{\tilde\mu}$, and observable $\calO$, we have
\begin{equation}
\label{chain}
\begin{gathered}
    |\langle\calO,\tilde\mu_b\rangle| =
    | \langle \calO, \calF^{-1}(\mu_b)\rangle|
    \le \|\calO\|_2 \|\calF^{-1}(\mu_b)\|_2 \\
    = \|\calO\|_2 \sqrt{\langle\mu_b,\calF^{-2}(\mu_b)\rangle}
    \le \|\calO\|_2 \| \calF^{-2}\|_{\rm op}^{1/2},
\end{gathered}
\end{equation}
where $\calF$ is the rescaled frame operator, $\|\cdot\|_{\rm op}$ is the operator norm, and $\|X\|_2\equiv \sqrt{\trace(X^\dagger X)}$ is the $L_2$ operator norm of $X$.
For the second identity we used the self-adjoint nature of the linear operator $\calF^{-1}$ to move it across the inner product, thus getting
\begin{equation}
    \|\calF^{-1}(\mu_b)\|_2^2 =
    \langle \calF^{-1}(\mu_b),\calF^{-1}(\mu_b)\rangle
    = \langle \mu_b,\calF^{-2}(\mu_b)\rangle.
\end{equation}
Moreover, we used the shorthand notation $\calF^{-2}\equiv \calF^{-1}\circ\calF^{-1}$.
The last step in the chain of relations in Eq.~\eqref{chain} then follows from
\begin{equation}
    \langle\mu_b,\mathcal F^{-2}(\mu_b)\rangle
    \le
    \|\mu_b\|_2^2 \|\mathcal F^{-2}\|_{\rm op} \le \|\mathcal F^{-2}\|_{\rm op}.
\end{equation}
As this holds for all $b$, $b\mapsto \hat o(b)$ is a bounded estimator.
This implies that Hoeffding-like performance guarantees can always be used, that is, that to have $\on{Pr}(|\overline o_N-\mathbb{E}[\hat o]|\ge\epsilon)\le\delta$, with $\overline o_N$ the sample mean taken over $N$ independently drawn samples, it is sufficient to use $N\ge \frac{C}{\epsilon^2}\log(2/\delta)$, with $C$ a constant independent from $\epsilon,\delta$.
This matches the type of performance guarantees provided by the median-of-means estimator, explaining why in many practical scenarios the standard mean can perform better than the median-of-means estimator.
Nonetheless, it is worth remarking that the constant $C$ will depend on the interval of values taken by the estimator $\hat o$, which as shown above are only upper bounded by $\|\mathcal F^{-2}\|_{\rm op}^{1/2}$. This quantity can increase with the state dimension $d$.
Consequently, while median-of-means is never useful from the perspective of the scaling of $N$ with respect to $\epsilon,\delta$, it might provide advantages in higher-dimensional spaces, as was found to be the case in the analytical derivation for Clifford circuits in~\cite{helsen2022thrifty}.
It is worth stressing that the results we present in this paper are completely agnostic to the choice of between means and median-of-means, as our analysis is performed at the level of the single-shot estimator. It is therefore entirely possible to apply the estimators we propose using either standard mean, median-of-means, or possibly estimators that provide even more advantageous bounds~\cite{minsker2023efficient}.

\section{Canonical estimators}
\label{sec:estimators_derivation}

\parTitle{Minimum-variance unbiased estimators for tomography}
It was shown~\cite{zhu2014quantum,zhu2011quantum,scott2006tight} in the context of state tomography that the operators $\tilde\mu_b^{{(\rho)}}$ defined as
\begin{equation}\label{eq:optimal_estimator_frame_op_tomography}
\labeltarget{cooleq:optimal_estimator_frame_op_tomography}
    \tilde\mu_b^{{(\rho)}} \equiv \frac{\calF_{\rho}^{-1}(\mu_b)}{\langle\mu_b,\rho\rangle},
    \qquad
    \calF_{\rho} \equiv \sum_b \frac{\mathbb{P}(\mu_b)}{\langle\mu_b,\rho\rangle},
\end{equation}
give an unbiased estimator that minimizes the $L_2$ state estimation error if the input state is $\rho$ and the measurement is $\bs\mu$.
Here $\calF_{\rho}$ is the frame superoperator associated to the rescaled measurement frame with elements $\mu_b/\sqrt{\langle\mu_b,\rho\rangle}$.
Note that $\tilde{\bs\mu}^{(\rho)}\equiv (\tilde\mu_b^{(\rho)})_b$ is a dual measurement frame for $\bs\mu$, but not its canonical dual measurement frame. It is a suitably rescaled version of the canonical dual to the rescaled measurement frame with elements $\mu_b/\sqrt{\langle\mu_b,\rho\rangle}$.
To use $\tilde{\bs\mu}^{(\rho)}$ one needs to already have a good guess about the underlying state $\rho$ which is being measured, and we thus interpret $\rho$ as the prior information on the input state~\footnote{Note that we use the term ``prior information'' informally, without introducing a prior distribution explicitly, since we do not adopt a Bayesian framework.
However, we mention that shadow tomography has recently been studied from a Bayesian perspective~\cite{lukens2020bayesian}.}.
Thus, $\tilde{\bs\mu}^{(\rho)}$ is the minimum-variance unbiased estimator when the input state is $\rho$.
A convenient tool to study the precision of an estimator is the \textit{MSE matrix}. Following~\cite{zhu2014quantum}, this is defined with respect to a generic dual frame $\tilde{\bs\mu}$ and state $\rho$ as
\begin{equation}\label{eq:mse_matrix_definition}
    \calC_{\rho} \equiv
    \sum_b \langle\mu_b,\rho\rangle \mathbb{P}(\tilde\mu_b) - \mathbb{P}(\rho).
\end{equation}
While we do not write the functional relationship explicitly, $\calC_\rho$ depends on the choice of $\bs\mu,\tilde{\bs\mu}$, and $\rho$.
The expected $L_2$ state estimation error associated to the estimator $\hat f(b)=\tilde\mu_b$ can be written concisely using the MSE matrix as
\begin{equation}
    \calE_\rho \equiv
    \mathbb{E}[\|\hat f-\rho\|_2^2] =
    \trace(\calC_\rho).
\end{equation}
When using the estimator $\tilde\mu_b=\tilde\mu_b^{(\rho)}$, the MSE matrix simplifies to
\begin{equation}
    \calE_\rho = \trace(\calF^{-1}_{\rho}) - \trace(\rho^2),
\end{equation}
which is the expected mean squared error when using the estimator with minimum-variance when the input state is $\rho$~\footnote{Note that these expressions can be written more rigorously as $\calE_\rho(\bs\mu,\tilde{\bs\mu}) = \trace(\calC_\rho(\bs\mu,\tilde{\bs\mu}))$ and $\calE_\rho(\bs\mu,\tilde{\bs\mu}^{(\rho)}) = \trace(\calF_\rho(\bs\mu)^{-1}) - \trace(\rho^2)$, making explicit the dependence of $\calF_\rho$ on the measurement $\bs\mu$.}.
In the expression $\trace(\calF^{-1})$, the argument $\calF^{-1}$ is a superoperator, but its trace is defined as in linear algebra for a standard trace. However, it is often the case that the trace of a superoperator is referred to as a ``superoperator trace''.
Explicitly, the (superoperator) trace of a generic superoperator $\Phi$ can be defined as $\trace(\Phi)=\sum_k \langle\sigma_k,\Phi(\sigma_k)\rangle$ for any orthonormal basis of operators $\{\sigma_k\}_k$.
In our case, $\calF_\rho^{-1}$ is considered as an operator acting in the subspace of Hermitian operators, and its trace is thus
\begin{equation}
    \trace(\calF_\rho^{-1}) = \sum_{k=1}^{d^2} \langle \sigma_k,\calF_\rho^{-1}(\sigma_k)\rangle
\end{equation}
with $\{\sigma_k\}_{k=1}^{d^2}$ a generic orthonormal basis of Hermitian operators, and $d$ the dimension of the underlying space.
It is also often convenient to pick an orthonormal basis of the form $\{I/\sqrt d\}\cup\{\tilde\sigma_k\}_{k=1}^{d^2-1}$, where $I/\sqrt d$ is the (normalized) identity, and $\{\tilde\sigma_k\}_{k=1}^{d^2-1}$ forms an orthonormal basis for the subspace of \textit{traceless} Hermitian operators. This can always be done, and is very useful in our calculations for a twofold reason. On one hand, it provides the following decomposition for the (superoperator) trace
\begin{equation}
    \trace(\calF_\rho^{-1}) = \frac{\trace(\calF_\rho^{-1}(I))}{d}
    + \sum_{k=1}^{d^2-1} \langle\tilde\sigma_k,\calF_\rho^{-1}(\tilde\sigma_k)\rangle.
\end{equation}
On the other hand, as $\calF^{-1}_\rho(I)=\rho$ --- which follows directly from the readily verifiable relation $\calF_\rho(\rho)=I$ --- we reduce the calculation of the trace to the calculation of the trace on the subspace of traceless Hermitian operators.

\parTitle{Canonical estimator}
A standard scenario is the lack of any prior information about the input state. In such cases, because the error will generally depend on the input state, it is common to consider as ``optimal'' the estimator that minimizes the \textit{average} $L_2$ estimation error, which corresponds to the optimal estimator with respect to the reference state $\rho=I/d$.
Following~\cite{zhu2014quantum}, we will refer to this as the \textit{canonical estimator}, denoted with $\tilde{\bs\mu}^{\rm can}\equiv\tilde{\bs\mu}^{(I/d)}$, which is thus written explicitly as
\begin{equation}\label{eq:definition_canonical_estimator}
\labeltarget{eq:definition_canonical_estimator}
    \tilde\mu_b^{\rm can} \equiv \frac{ d\calF_{I/d}^{-1}(\mu_b)}{ \trace(\mu_b) }, \qquad
    \calF_{I/d}\equiv d\sum_b \frac{\mathbb{P}(\mu_b)}{\trace(\mu_b)}.
\end{equation}
It is worth noting that this is not the same as the canonical dual with respect to the measurement frame $\bs\mu$~\footnote{The notational mismatch is analogous to how a ``tight measurement frame'' is not a ``tight frame'' built from the POVM elements. These differences can be traced back to the fact that any POVM satisfies the normalization condition $\sum_b \mu_b=I$, which imposes some structure on the associated frame operator and derived quantities, which is easier to handle introducing the specific notions of ``tight measurement frame''. On the other hand, the mismatch in the use of the term ``canonical dual'' is due to the fact that if the goal is minimizing the variance corresponding to the estimator, the optimal choice is using the canonical dual (in the standard sense) of a \textit{rescaled} measurement frame. These aspects are discussed more in-depth in~\cref{sec:rescaled_frames}.}.
The canonical estimator thus minimizes the $L_2$ error averaged over unitarily equivalent input states~\cite{scott2006tight,roy2007weighted,zhu2011quantum}.
This average $L_2$ error turns out to only depend on the purity $P\equiv\trace(\rho^2)$ of the input state, and will be denoted with $\overline{\calE_P}$. As discussed in Ref.~\cite{zhu2014quantum}, this quantity is lower bounded by
\begin{equation}
    \overline{\calE_P} \ge 
    d^2 + d - 1 - P,
\end{equation}
with the lower bound saturated iff the measurement is composed of projectors onto subnormalized pure states that form a weighted 2-design.
Such measurements are referred to as \textit{tight rank-1 IC-POVMs}, and have elements $\mu_b=w_b \mathbb{P}(\psi_b)$ with the weights satisfying $\sum_b w_b=d$, and
\begin{equation}
    \frac1d\sum_b w_b \mathbb{P}(\psi_b)^{\otimes 2} = {\binom{d+1}{2} }^{-1}\Pi_{\rm sym},
\end{equation}
with $\Pi_{\rm sym}$ the projection onto the symmetric subspace, that can be written explicitly as $\Pi_{\rm sym}=(I+W)/2$, with $W$ the Swap operator.
For all tight rank-1 IC-POVMs, the canonical estimator has the form
\begin{equation}\label{eq:canonical_estimator_tight}
    \tilde\mu_b^{\rm can} = 
    (d+1) \mathbb{P}(\psi_b) - I.
\end{equation}
and the MSE matrix equals
\begin{equation}
    \calC_{I/d} = \frac{d+1}{d} \Pi_{H_0},
\end{equation}
where $\Pi_{H_0}\equiv \on{Id} - \mathbb{P}(I/\sqrt d)$ is the superoperator that projects onto the subspace of traceless linear operators.
A more in-depth discussion of these results, and more generally of the connection between weighted 2-designs and tight IC-POVMs, is given in~\cref{app:sec:tightmeasurements_vs_weighted2designs}.

\parTitle{Estimation of observables}
The usefulness of shadow tomography lies in the potentially favorable scalings of the associated estimation errors with respect to the state dimension $d$.
More specifically, we are interested in the variance of $\hat o$ for different choices of $\rho,\bs\mu,\tilde{\bs\mu}$, and $\calO$. For notational convenience, we indicate explicitly only the dependence of the variance on $\rho$:
\begin{equation}\label{eq:general_variance}\small
    \on{Var}[\hat o|\rho] 
    =\mathbb{E}[|\hat o - \langle \calO,\rho \rangle |^2 ]
    = \sum_b \langle\mu_b,\rho\rangle \langle\calO,\tilde\mu_b\rangle^2 -
    \langle\calO,\rho\rangle^2
\end{equation}
for different choices of $\rho,\bs\mu,\tilde{\bs\mu},\calO$.
This can be conveniently written using the MSE matrix $\calC_\rho$ as
\begin{equation}\label{eq:variance_observable_estimator}
    \on{Var}[\hat o|\rho] = \langle \mathbb{P}(\calO), \calC_\rho\rangle \equiv
    \langle\calO,\calC_\rho(\calO)\rangle.
\end{equation}

As discussed in detail in~\cref{sec:optimal_estimator_observables}, we can derive a general expression for the minimum-variance unbiased estimator for a given target observable and input state, and this is found to match the corresponding estimator for state tomography on the support of the observable.
More precisely, if $\tilde{\bs\mu}^{(\rho)}$ is a minimum-variance unbiased estimator for state tomography with respect to the state $\rho$, then any $\tilde{\bs\mu}$ such that $\langle\calO,\tilde\mu_b\rangle=\langle\calO,\tilde\mu_b^{(\rho)}\rangle$ is a minimum-variance unbiased estimator for $\calO$.
Although derived using different methods and notation, this result is similar to some of the results reported in~\cite{perinotti2007optimal,dariano2007optimal,tran2022measuring}. If we want an estimator which gives small errors for arbitrary target observables, the natural candidate is to use the one minimizing the variance averaged over the observables.
In this case, the minimum-variance unbiased estimator is again the one we found for state tomography.
Given that in shadow tomography we do not generally want to fix beforehand the observables to estimate, we can safely fix as optimal estimators the $\tilde{\bs\mu}^{(\rho)}$ derived for state tomography.
We will furthermore focus on the scenario where only the purity of the input state is known beforehand, and we will thus in the following always use the canonical estimator $\tilde{\bs\mu}^{\rm can}$ given in~\cref{eq:definition_canonical_estimator}.
This has the added advantage of being independent of both $\rho$ and $\calO$, though the estimation variance will still in general depend on these quantities.
In summary, if there is prior information suggesting that the input state is or is close to $\rho$, the minimum-variance estimator is given by $\tilde{\bs\mu}^{(\rho)}$, as discussed in this section, and proved explicitly in~\cref{sec:optimal_estimator_state,sec:optimal_estimator_observables}. If no prior knowledge is assumed about the input state, the canonical estimator $\tilde{\bs\mu}^{\rm can}$ can be used and provides the minimal averaged estimation variance.

\begin{figure}[tb]
    \centering
    \includegraphics[width=0.45\textwidth]{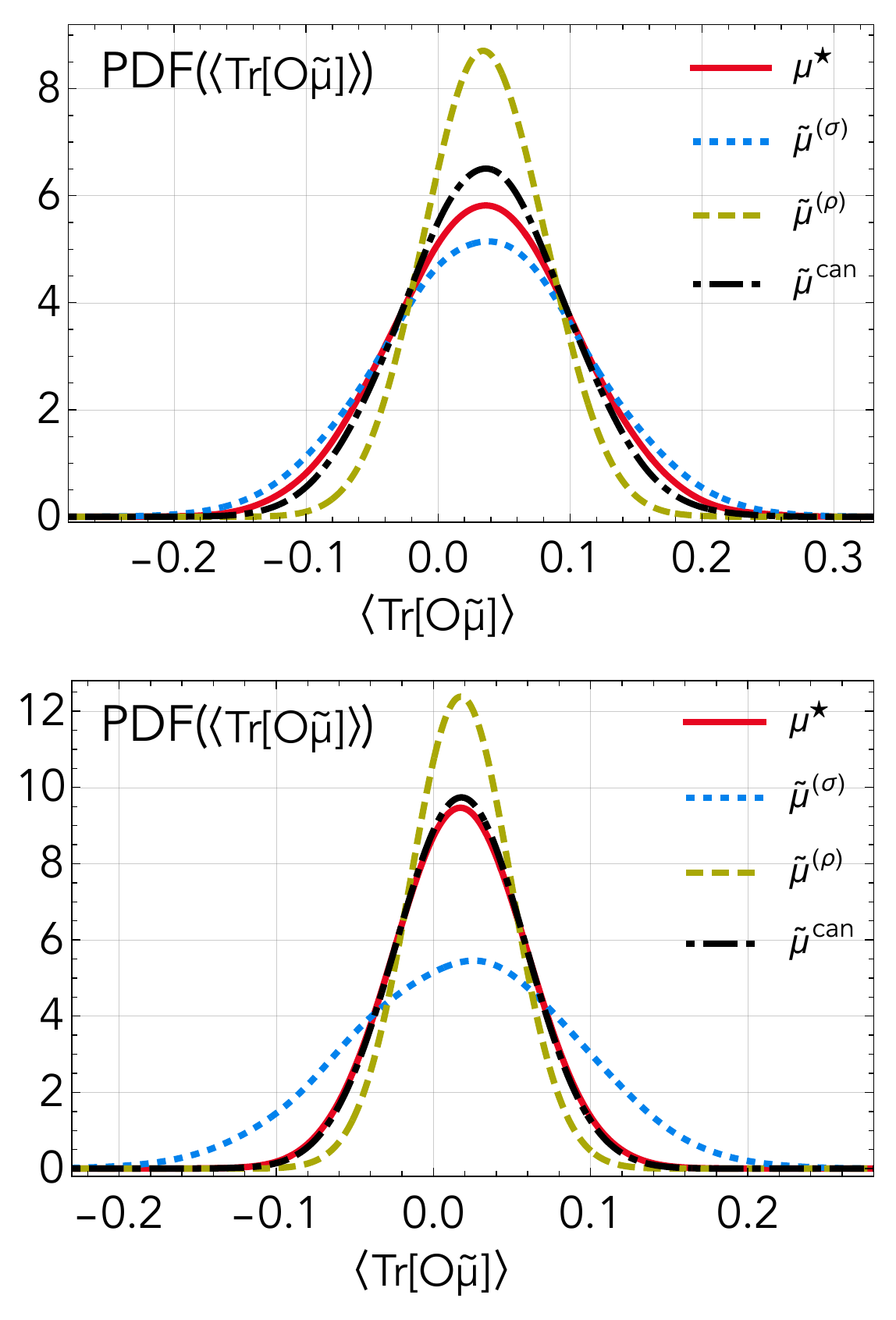}
    \caption{
    \textbf{Probability distributions of samples means.}
    Histograms of the probability distribution of the sample mean $\overline o_N$ with $N=10^3$, obtained taking the average of $\hat o(b)\equiv \langle \hat f(b),\mathcal O\rangle$ over $N$ randomly sampled outcomes $b$, for different choices of estimator $\hat f$.
    The histograms are computed using $10^4$ realizations of the sample mean.
    The input state is $\rho\equiv\mathbb{P}_0$ in all cases, and the measurements are random rank-1 POVMs built as $\mu_b=V \mathbb{P}_b V^\dagger$ with $V$ random isometries.
    In each case we show the distribution of the sample mean for (1) the non-rescaled estimator $\mu^\star$ (\textit{c.f.}~\cref{eq:frame_operator_def}); (2, 3) the estimators  $\tilde\mu^{(\rho)}$ and $\tilde\mu^{(\sigma)}$ (\textit{c.f.}~\cref{eq:optimal_estimator_frame_op_tomography})
    with $\sigma\equiv\mathbb{P}_1$;
    (4) the canonical estimator $\tilde\mu^{\rm can}$ (\textit{c.f.}~\cref{eq:definition_canonical_estimator}).
    We show the data for 
    \textbf{(up)}
    $2$-dimensional states with $10$-outcome measurements, and 
    \textbf{(down)}
    $5$-dimensional states and $100$-outcome measurements.
    }
    \label{fig:histograms_obs_estimators}
\end{figure}

\begin{figure}[tb]
    \centering
    \includegraphics[width=0.45\textwidth]{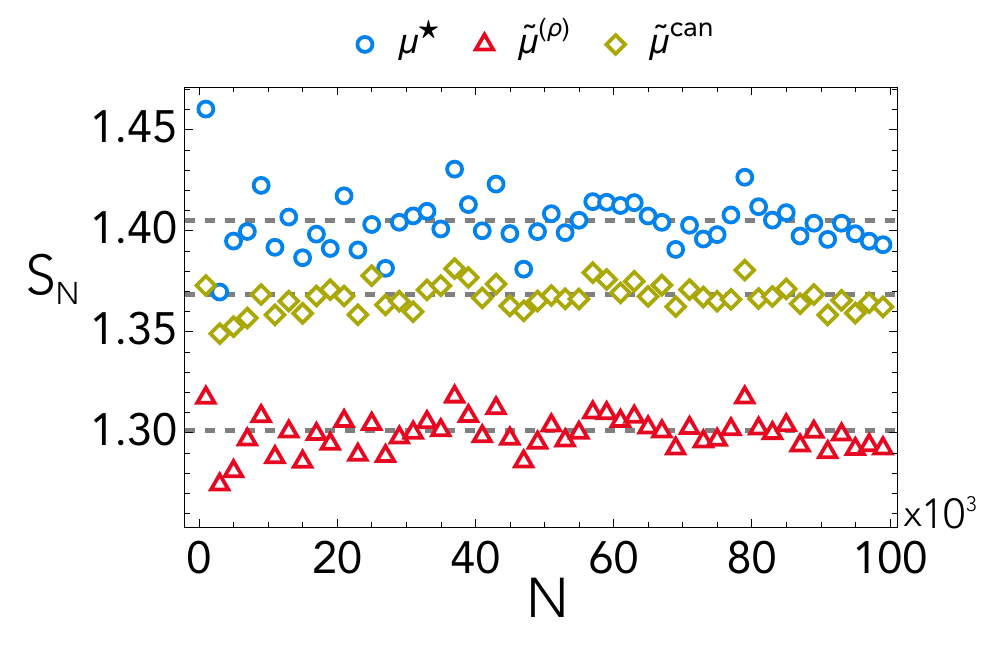}
    \caption{
    \textbf{Sample variance for different estimators.}
    Examples of behaviour of the sample variance $\hat S_N$ of the estimator $\overline o_N\equiv\frac1N\sum_{k=1}^N \hat o(b_k)$ as a function of $N$, computed with respect to the estimators $\bs\mu^\star$, $\tilde{\bs\mu}^{(\rho)}$, and $\tilde{\bs\mu}^{\rm can}$.
    The sample variance is defined as
    $\hat S_N \equiv \frac{1}{N-1}\sum_{k=1}^{N}(\hat o(b_k)-\overline o_N)^2$.
    The dashed lines give the values of the variance $\on{Var}[\hat o|\rho]$ in each case, as computed via~\cref{eq:general_variance}.
    The data is obtained using $d=2$-dimensional systems, with fixed input state $\rho=\mathbb{P}_0$, random rank-1 POVMs with 10 outcomes, and random target observables with $\trace(\calO)=0$ and $\trace(\calO^2)=1$.
    }
    \label{fig:sample_variance}
\end{figure}

\parTitle{Numerical examples}
We illustrate explicitly how different choices of dual frames provide non-equivalent estimators in~\cref{fig:sample_variance,fig:histograms_obs_estimators}.
In particular, the canonical estimator $\tilde{\bs\mu}^{\rm can}$ has on average the lowest variance, albeit the estimator $\tilde{\bs\mu}^{(\rho)}$ can give even lower variances if $\rho$ matches the true input state. The nonrescaled estimator $\bs\mu^\star$ tends to perform worse than $\tilde{\bs\mu}^{\rm can}$, consistently with the latter having a smaller averaged variance. On the other hand, using the estimator $\tilde{\bs\mu}^{(\sigma)}$ --- which has minimum variance when the input is $\sigma$ --- to estimate properties of $\rho\neq\sigma$, will still reproduce on average the correct expectation values, but result in a generally larger estimation error.

\parTitle{Shadow tomography vs state tomography}
It is worth stressing the tight relation between shadow and state tomography emerging from the above discussion.
The general formalism of measurement frames clarifies how these can be viewed as one and the same experimental protocol, with the only difference being how estimation errors are evaluated.
Both linear state tomography and our formalism for shadow tomography can be performed for arbitrary IC-POVMs --- albeit, as discussed previously, not always with favorable error scalings --- and the post-processing procedure is the same in both cases. The core difference is in the problem setting: whether the target is recovering an approximation of the full density matrix, or just recovering the expectation values of finitely many observables.

\section{Bounds on averaged variance}
\label{sec:bounds_averaged_variance}

In this section, we will derive useful bounds for the averaged estimation variance of an observable in terms of the eigenvalues of the frame superoperator associated to the measurement. These eigenvalues will then be bounded in terms of a quantity that measures how far a given IC-POVM is from being tight. Finally, we will show that, for tight measurements and any suitable normalized observable, the resources needed to estimate the expectation value of the observable via the shadow tomography apparatus do not scale with the dimension of the state.

\parTitle{Bounds via eigenvalues of frame superoperator}
The variance averaged over unitarily equivalent input states is
\begin{equation}\label{eq:averaged_variance_general_formula}
\begin{gathered}
    \overline{\on{Var}[\hat o|P,\calO,\bs\mu]}
    \equiv \int_{\mathbf U(d)} dU\, \on{Var}[\hat o|U\rho U^\dagger] \\
    = \underbrace{
    \sum_b \langle \mu_b,I/d\rangle \langle\calO,\tilde\mu_b^{\rm can}\rangle^2
    }_{=\langle\calO,\calF_{I/d}^{-1}(\calO)\rangle} - \underbrace{
    \int_{\mathbf U(d)}dU \langle\calO,U\rho U^\dagger\rangle^2
    }_{\equiv\beta}.
\end{gathered}
\end{equation}
As mentioned previously, the explicit dependence on $\calO$ and $\bs\mu$ will be left implicit in the following for notational conciseness, and we will write this averaged variance as simply $\overline{\on{Var}[\hat o|P]}$.
The coefficient $\beta$, whose explicit expression is reported in~\cref{sec:errors:singleobservables}, is computed explicitly using known formulas to integrate polynomials in the components of unitaries matrices over the uniform Haar measure~\cite{collins2003moments,collins2022weingarten}, and does not depend on $\bs\mu$. Furthermore, as shown in~\Cref{sec:properties_frame_ops}, the canonical superoperator decomposes as
\begin{equation}
    \calF_{I/d} = d \mathbb{P}(I/\sqrt d)+\tilde\calF_{I/d}
\end{equation}
with $\tilde\calF_{I/d}\equiv \Pi_{H_0}\calF_{I/d}\Pi_{H_0}$ the projection of $\calF_{I/d}$ onto the subspace of traceless operators. Using such decomposition, we rewrite
\begin{equation}
    \langle\calO,\calF_{I/d}^{-1}(\calO)\rangle =
    \frac{\trace(\calO)^2}{d^2} +
    \langle\calO,\tilde\calF_{I/d}^{-1}(\calO)\rangle.
\end{equation}
The second term can then be bounded in terms of the eigenvalues of $\tilde\calF_{I/d}$, as
\begin{equation}\label{eq:averaged_variance_vs_eigenvalues}
    \frac{Vd}{ \lambda_{+}(\tilde\calF_{I/d}) }
    \le 
    \langle\calO,\tilde\calF_{I/d}^{-1}(\calO)\rangle
    \le
    \frac{Vd}{ \lambda_{-}(\tilde\calF_{I/d}) },
\end{equation}
where $\lambda_{-}(\tilde\calF_{I/d})$, $\lambda_{+}(\tilde\calF_{I/d})$ denote the smallest and largest eigenvalues of $\tilde\calF_{I/d}$, respectively, and $V\equiv \trace(\calO^2)/d-\trace(\calO)^2/d^2$ is the variance of $\calO$ with respect to the totally mixed state $I/d$.
As further explained in~\cref{sec:errors:singleobservables}, this expression is obtained observing that $\tilde\calF_{I/d}^{-1}$ is a Hermitian linear (super)operator which only acts nontrivially on the subspace of traceless Hermitian operators.
Being $\tilde\calF_{I/d}$ positive definite as an operator whenever $\bs\mu$ is informationally complete, we are ensured that $\lambda_\pm(\tilde\calF_{I/d})>0$.
For any $\bs\mu$, as again showed in~\cref{sec:errors:singleobservables}, the eigenvalues can be bounded as a function of $a\equiv \trace(\tilde\calF_{I/d})$ and $b\equiv \trace(\tilde\calF_{I/d}^2)$.
Focusing on the variance for the hardest-to-estimate observable, we find that the smallest such variance compatible with $a,b$ reads
\begin{equation}\label{eq:worst-case-variance-bound}
\begin{gathered}
    \max_{\calO} \frac{
        \overline{
            \on{Var}[\hat o|P]
        }
    }{
        Vd
    } \ge \frac{1}{\lambda_1^*} - \frac{P-1/d}{d^2-1}, \\
    \lambda_1^* \equiv \frac{a}{d^2-1}
    - \frac{\sqrt{(d^2-2)((d^2-1)b-a^2)}}{(d^2-1)(d^2-2)}.
\end{gathered}
\end{equation}
This relation tells us that if $\bs\mu$ gives a frame superoperator such that $a=\trace(\tilde\calF_{I/d})$ and $b=\trace(\tilde\calF_{I/d}^2)$, then the worst-case average variance is lower bounded as in~\cref{eq:worst-case-variance-bound}.
In other words, $a$ and $b$ define a bound on the best possible performance of the canonical estimator (in the scenario where we average over input states and take the worst-case scenario with respect to observables).

\parTitle{Performances for tight measurements}
In the case of tight measurements, $\tilde\calF_{I/d}$ is a multiple of the identity, $(d^2-1)b=a^2$, and~\cref{eq:worst-case-variance-bound} simplifies to
\begin{equation}\label{eq:conclusion_upperbound_obsestimationerror}
    \overline{\on{Var}[\hat o|P]}
    = Vd \left(\frac{d^2+d-1-P}{d^2-1}\right),
\end{equation}
where the max over observables does not apply anymore because all observables give the same expression for the averaged average.
We recognize in particular the term $d^2+d-1-P$ which is the optimal state estimation $L_2$ error discussed in~\cref{app:sec:tightmeasurements_vs_weighted2designs}.
\Cref{eq:conclusion_upperbound_obsestimationerror} shows that for tight rank-1 measurements, the variance increases with the state dimension only due to the variance $V$ of the observable calculated with respect to the totally mixed state.
Note that for rank-1 observables of the form $\calO=\mathbb{P}_\psi$ for any $\ket\psi$, we have $Vd=1-1/d$, while for observables normalized as $\trace(\calO)=0$ and $\trace(\calO^2)=1$, we have $Vd=1$.
It immediately follows that for all such cases $Vd\to1$ for large $d$, and thus the variance does not increase with $d$, converging asymptotically to $V\to1$.
On top of estimating best- and worst-case scenarios for the variance, we show in~\cref{sec:error_avg_observables} how to also compute it averaging with respect to unitarily equivalent observables.

We thus showed that for the entire class of tight rank-1 measurement frames, which includes but is not limited to covariant measurements, the sampling statistics required to estimate arbitrary rank-1 observables with bounded norm does not increase with the state dimension, in direct contrast with the corresponding results about state tomography.
More generally, we can explicitly characterize the class of observables which correspond to such favorable scalings.
This directly implies that all these measurements can be used to implement shadow tomography schemes.
While not all such measurements will allow an efficient circuit decomposition like the one presented in~\cite{huang2020predicting}, this will depend on the experimental context that is being considered. Having a good characterization of the general class of viable measurements can greatly help to find measurement schemes to efficiently implement shadow tomography in different experimental scenarios.

\section{Best- and worst-case scenario variances}
\label{sec:minmax_variance}

In~\cref{sec:bounds_averaged_variance} we derived bounds for the variance averaged over input states. In this section, we focus instead on the derivation of bounds for minimum and maximum variance with respect to the input states. This is particularly relevant for comparing with the results of~\cite{huang2020predicting} because, as will be discussed in detail in~\cref{sec:original_formalism}, the often used ``shadow norm'' is precisely the variance maximized over input states.

\parTitle{Concise expression for variance via $A$ operator}
We first observe that the general expression for the variance in~\cref{eq:general_variance}, for a generic input state $\rho$, can be rewritten as
\begin{equation}\label{eq:variance_vs_Arho}
    \on{Var}[\hat o|\rho] + \langle\calO,\rho\rangle^2 =
    \sum_b \langle\mu_b,\rho\rangle 
    \langle\calO,\tilde\mu_b\rangle^2
    = \langle A,\rho\rangle,
\end{equation}
where we defined the operator
\begin{equation}
\label{eq:definition_A_op}
    A\equiv\sum_b \langle\calO,\tilde\mu_b\rangle^2\mu_b.
\end{equation}
Notably, the only part of~\cref{eq:variance_vs_Arho} nonlinear with respect to $\rho$ is $\langle\calO,\rho\rangle^2$, which does not depend on the measurement choice, and is bounded as
$
    \langle\calO,\rho\rangle^2 \le \trace(\calO^2).
$ 
Furthermore, the linearity of $\langle A,\rho\rangle$ with respect to $\rho$ means that for any choice of measurement, estimator, and observable, we can write the general bounds:
\begin{equation}\label{eq:Arho_bound_via_eigenvalues}
    \lambda_{\rm min}(A) \le
    \langle A,\rho\rangle \le
    \lambda_{\rm max}(A) \equiv \|A\|_{\rm op},
\end{equation}
where $\lambda_{\rm min}(A), \lambda_{\rm max}(A)$ are smallest and largest eigenvalues of $A$, respectively.
In particular, we have the following upper bound for the worst-case (with respect to input states) variance:
\begin{equation}
    \max_\rho\on{Var}[\hat o|\rho] \le \|A\|_{\rm op}.
\end{equation}
As will be further discussed in more detail in~\cref{sec:original_formalism}, the right-hand side of this expression corresponds to the so-called ``shadow norm'' $\|\calO\|_{\rm sh}^2=\|A\|_{\rm op}$ introduced in Ref.~\cite{huang2020predicting}.

\begin{figure}[ht]
\centering
\includegraphics[width=0.35\textwidth]{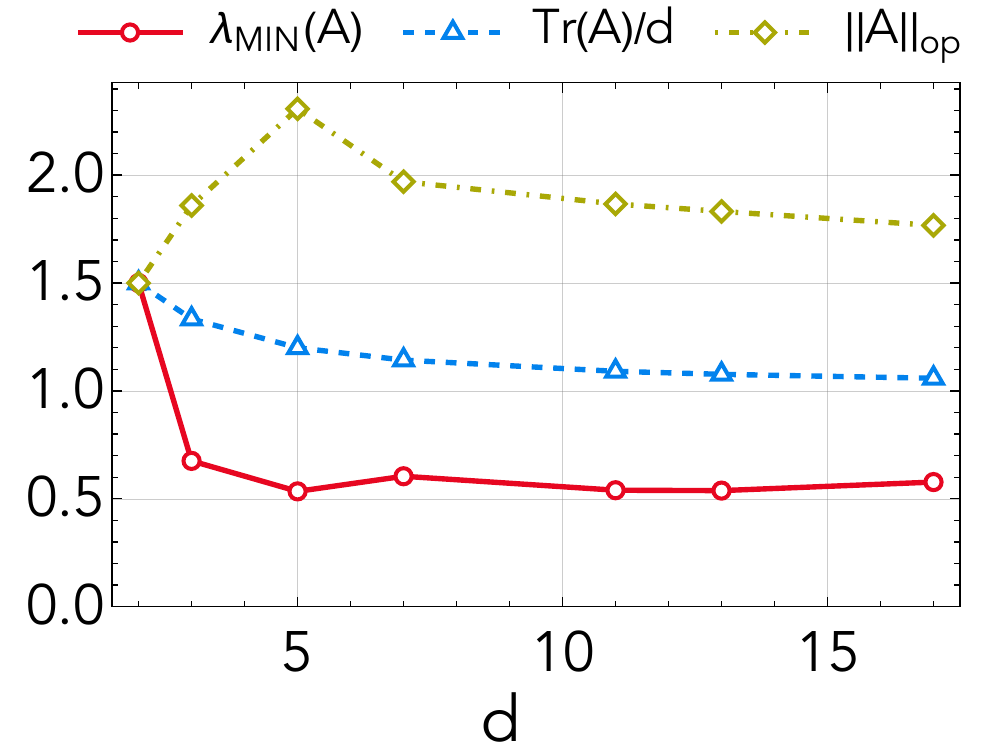}
\caption{
\textbf{Average, min, and max variance for MUB POVMs.}
We plot the values of $\lambda_{\rm min}(A)$, $\|A\|_{\rm op}$, and $\trace(A)/d$, as a function of the state dimension $d$, for the case of canonical estimators, with a random target observable for each $d$.
The data is shown for prime $d$ because these are the values corresponding to which explicit constructions for MUBs are known~\cite{durt2010mutually}.
These results give the range of possible values of $\langle A,\rho\rangle$ varying over the input states $\rho$, for the case of MUB measurements. These values are then tightly connected with the estimation variance via~\cref{eq:variance_vs_Arho}.
The data shown corresponds to a random target observable with $\trace(\calO)=0$ and $\trace(\calO^2)=1$.}
\label{fig:avg_min_max_variance_mubs}
\end{figure}

\parTitle{Explicit expression for 3-designs}
In the case of rank-1 measurements that also give a weighted 3-design, we can find a remarkably simple expression for the state- and observable-dependent variance even in the non-averaged scenario.
To see this, we start observing that
\begin{equation}\small
    \sum_b \langle\mu_b,\rho\rangle
    \langle\calO,\tilde\mu_b\rangle^2  =
    \left\langle \rho\otimes\calO\otimes\calO,
    \sum_b \mu_b\otimes\tilde\mu_b\otimes\tilde\mu_b \right\rangle.
    \label{eq:variance_3design_tensor_products}
\end{equation}
For any tight rank-1 POVM with elements $\mu_b=w_b \mathbb{P}(\psi_b)$, using the canonical estimators $\tilde\mu_b^{\rm can}$ given in~\cref{eq:canonical_estimator_tight}, we can also write
\begin{equation}
    \sum_b \mu_b\otimes\tilde\mu_b^{\rm can}\otimes \tilde\mu_b^{\rm can}
    = (d+1)^2 S_3
    - (d+1) S_2 + I,
\end{equation}
where $S_3 \equiv \sum_b w_b\mathbb{P}(\psi_b)^{\otimes3}$ and 
\begin{equation}
    S_2\equiv
    \sum_b w_b\mathbb{P}(\psi_b)^{\otimes2}\otimes I
    +
    \sum_b w_b\mathbb{P}(\psi_b)\otimes I\otimes \mathbb{P}(\psi_b).
\end{equation}
If the states $\ket{\psi_b}$ form a complex projective 3-design, then
$S_3= d\Pi_{\rm sym,3}/\binom{d+2}{3}$ with $\Pi_{\rm sym,3}\in\on{Lin}((\mathbb{C}^d)^{\otimes3})$ the projection onto the completely symmetric subspace of $(\mathbb{C}^d)^{\otimes3}$, and $S_2\binom{d+1}{2}=d\Pi_{\rm sym,2}^{(1,2)}+d\Pi_{\rm sym,2}^{(1,3)}$ is a sum of the projections on the symmetric subspace of $(\mathbb{C}^d)^{\otimes2}$ on first and second and first and third qubits, respectively.
These projections can be written more explicitly as $\Pi_{\rm sym,2}=(I\otimes I+W)/2$ with $W$ the Swap operator,
$\Pi_{\rm sym,3}=\frac1{3!}\sum_{\pi\in \mathcal{S}_3 } W_\pi$ with $\mathcal{S}_3$ denoting the symmetric group over 3 elements, and $W_\pi$ the unitary operator defined as~\cite{watrous2018theory}
\begin{equation}
    W_\pi=\sum_{i_1,i_2,i_3}
    \ketbra{i_{\pi(1)}, i_{\pi(2)}, i_{\pi(3)} }{i_1, i_2,i_3}.
\end{equation}
With these and~\cref{eq:variance_3design_tensor_products} we can work out the explicit expressions for state- and observable-dependent variances, and obtain
\begin{equation}\label{eq:nonaveraged_var}\small
\begin{aligned}
    \on{Var}[\hat o|\rho]
    =
    &- \frac{\trace(\calO)^2 + 2\trace(\calO)\trace(\rho\calO)}{d+2}
    \\
    &+\frac{d+1}{d+2}\left[
    \trace(\calO^2) + 2\trace(\calO^2\rho)
    \right]
    - \trace(\calO\rho)^2.
\end{aligned}
\end{equation}
This expression shows explicitly that for any rank-1 measurement that forms a 3-design, we get an explicit expression for the variance even in the non-averaged regime.
This dramatically simplifies the study of the relations between best, worst, and average cases with respect to both input state and target observable.
Random Clifford circuits and Haar-random unitaries, considered in Ref.~\cite{huang2020predicting}, as well as single-qubit mutually unbiased bases, are examples of rank-1 measurements that form a 3-design~\cite{kueng2015qubit,webb2015clifford}.

\parTitle{Worst-case variance bounds for 3-designs}
The explicit expression for the variance for 3-designs allows to also derive general bounds for the variance maximized over the input states: given any rescaled observable, $\trace(\calO)=0$, we get from~\cref{eq:nonaveraged_var}:
\begin{equation}
    \max_\rho \on{Var}[\hat o|\rho]\le
    \trace(\calO^2) + 2\|\calO^2\|_{\rm op}
    \leq 3  \trace(\calO^2),
\end{equation}
which shows that increasing the dimension $d$, even in the worst-case scenario, the variance only increases with $d$ via the observable. Thus for any rescaled observable for which $\trace(\calO)=0$, $\trace(\calO^2)=1$, we get a dimension-independent upper bound.
Note that the $3\trace(\calO^2)$ upper bound is identical to the one derived in~\cite{huang2020predicting} for random Clifford and unitary measurements.

\parTitle{Numerical examples with MUBs}
In~\cref{fig:avg_min_max_variance_mubs} we report numerical results obtained for average, min, and max variance, in the case of MUB measurements in prime dimensions~\cite{durt2010mutually}, calculated via~\cref{eq:Arho_bound_via_eigenvalues}. We note in particular how even the worst-case variance does not increase with the state dimension. This is compatible with the general expression for the variance we will obtain for 3-designs, although MUBs do not correspond to a 3-design, indicating these favourable scaling results might hold even more generally.

\section{Relation with construction of Ref. [13]}

\label{sec:original_formalism}

We now specialize our discussion in~\cref{sec:shadow_tomography_formalism} to the formalism presented in Ref.~\cite{huang2020predicting}.
The goal is to show that the latter can be viewed and studied from the general perspective of measurement frames, and corresponds to the special case where the employed IC-POVM is a \textit{covariant measurement}~\cite{dariano2004informationally,zhu2014quantum,matthews2009distinguishability,zhu2012phdthesis}.

\parTitle{Description of the formalism}
The procedure to build classical shadows introduced in Ref.~\cite{huang2020predicting} involves the following steps
\begin{enumerate}
    \item Perform a random unitary rotation $\rho\mapsto U\rho U^\dagger$ on the state, and then measure the evolved state in the computational basis $\ket b$.
    \item Define the operator
    \begin{equation}\label{eq:definition_M_op}
    \begin{gathered}
        \mathcal M(\rho) \equiv \mathbb{E}\left[U^\dagger|\hat b\rangle\!\langle \hat b| U\right] \\
        \equiv
        \mathbb{E}_{U\sim\mathcal U}\sum_b
        \expval{U\rho U^\dagger}{b}\,
        U^\dagger\ketbra{b}{b} U,
    \end{gathered}
    \end{equation}
    where $\vert{\hat b}\rangle$ is a random variable associating to each outcome $b$ the corresponding state $\ket b$.
    The expectation value is taken with respect to some distribution $\calU$ in the group of unitary matrices, and with respect to the possible outcomes $b$ for each choice of unitary.
    \item Compute and store the operators $\hat\rho\equiv \mathcal M^{-1}(U^\dagger|\hat b\rangle\!\langle \hat b| U)$. These are referred to as the ``classical shadows'' of the state.
\end{enumerate}
To estimate the expectation values of an observable $\calO$, one then uses the estimator $\hat o\equiv \langle\calO,\hat\rho\rangle$ built from the classical shadows.
We will focus here on the task of estimating expectation values, although in~\cite{huang2020predicting} the estimation of other kinds of quantities is also discussed.
Another important aspect discussed in~\cite{huang2020predicting} is the efficiency of computing and storing the classical shadows for large many-qubit Hilbert spaces, which can be solved by leveraging Clifford circuits and the formalism of stabilizer states. We will not focus on these aspects here, but rather on the general structure of shadow tomography protocol.

\parTitle{Equivalence: step 1}
The equivalence between the formalism thus outlined and our approach is seen observing that a measurement in the computational basis $\{|b\rangle\}$ after evolving the state through a random unitary rotation $U$, amounts to a direct measurement with the POVM having elements
\begin{equation}
    \mu_{U,b} \equiv U^\dagger \ketbra{b}{b} U.
\end{equation}
\begin{figure}[ht]
\centering
    \includegraphics[width=0.45\textwidth]{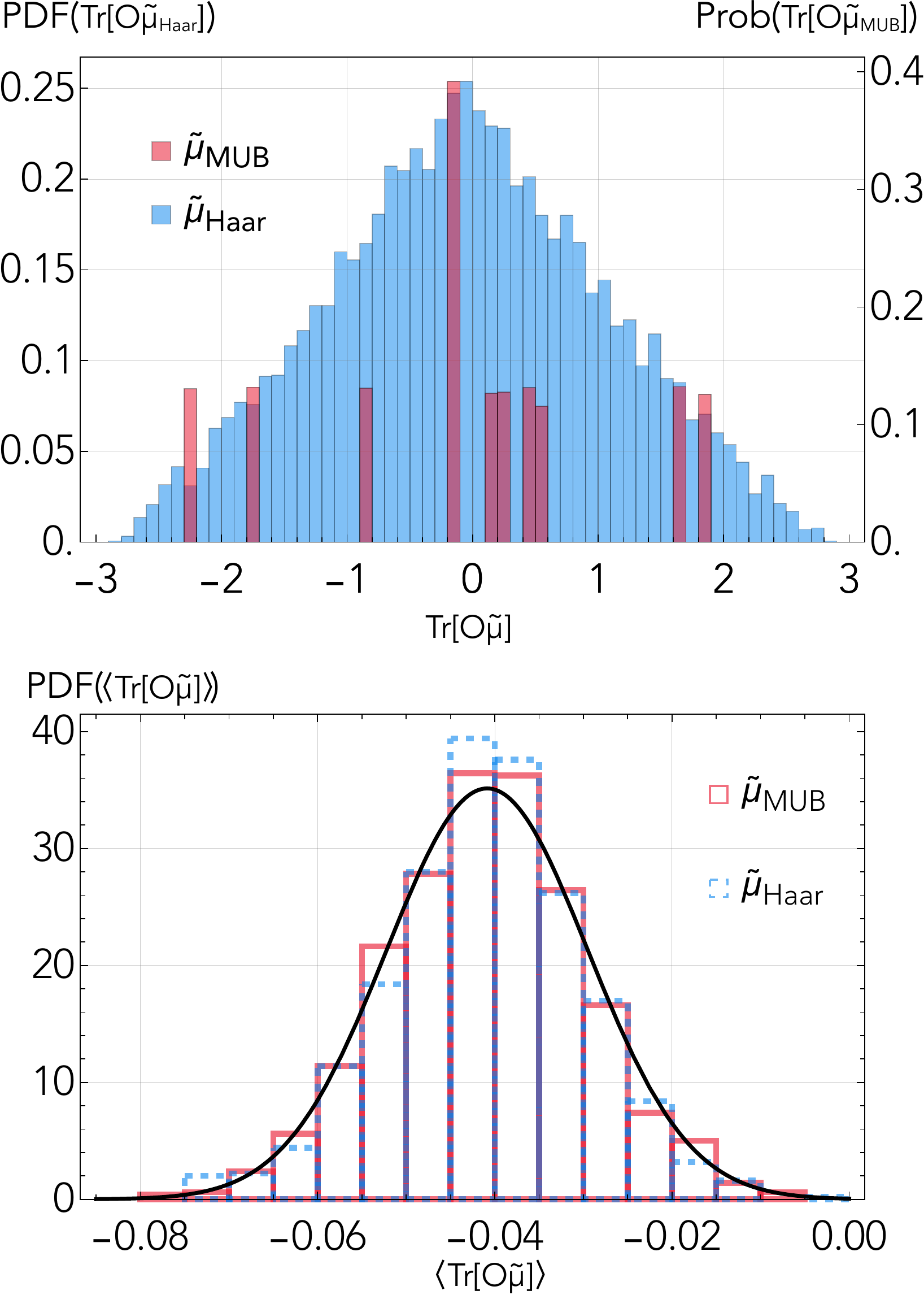}
    \caption{
    \textbf{Distributions of estimators and their sample mean, corresponding to MUBs and Haar-random unitary POVMs.}
    \textbf{(up)}
    Histograms of the probability distributions for the estimator $\langle\calO,\tilde\mu\rangle$ for a random (fixed) observable $\calO$ with $\trace(\calO)=0, \trace(\calO^2)=1$, and fixed qutrit state $\rho=\mathbb{P}_0$.
    The reported results correspond to MUBs, $\bs\mu_{\rm MUB}$ (red), and random measurements $\bs\mu_{\rm Haar}$, which have elements $\mu_{U,b}=U\mathbb{P}_b U^\dagger$ with Haar-random unitaries $U$ (blue).
    For $\bs\mu_{\rm MUB}$ there is a finite number of outcomes, and we directly plot the probability associated to each outcome.
    For $\bs\mu_{\rm Haar}$, owing to the infinitely many outcomes, we uniformly draw a number of random unitaries $U$, and plot a histogram of the observed estimator values $\langle\tilde{\bs\mu}_{\rm Haar},\calO\rangle$.
    We show two different scales on the vertical axis: in the presence of a continuum of possible outcomes, as we have for $\langle \calO,\tilde{\bs\mu}_{\rm Haar}\rangle$, we plot the probability density function (PDF), while for finitely many outcomes we show the probability mass function.
    \textbf{(down)} Histogram of possible outcomes of the sample mean $\overline o_N\equiv \frac 1N\sum_{k=1}^N \hat o(b_k)$ of $\hat o(b)\equiv \langle\calO,\tilde\mu_b\rangle$, estimated with a statistics of $N=10^3$ samples. The histogram is drawn sampling $10^4$ realizations of this sample mean, in the same condition as the other histogram.
    The black solid line is a Gaussian with the same mean and variance as both estimators $\tilde{\bs\mu}_{\rm MUB}$, $\tilde{\bs\mu}_{\rm Haar}$ --- which have the same variance, being both tight measurement frames. Both histograms approach this Gaussian for $N\to\infty$, due to the central limit theorem.
    }
\label{fig:estimator_values_probabilities}
\end{figure}
As such measurement has (uncountably) infinitely many outcomes, its normalization reads
\begin{equation}
    \int_{\mathbf U(d)} dU \sum_b \mu_{U,b} = I,
\end{equation}
where the integral is performed with respect to the Haar measure over the unitary group of suitable dimension, and thus $\int_{\mathbf U(d)}dU=1$.

\parTitle{Equivalence: step 2}
The introduced map $\mathcal M$ is precisely the frame operator corresponding to the measurement frame $\{\mu_{U,b}\}_{U,b}$. This becomes more evident rewriting~\cref{eq:definition_M_op} in the form:
\begin{equation}\label{eq:definition_M_op_explicit}
    \mathcal M(\rho) = \int_{\mathbf U(d)} dU \sum_b \langle\mu_{U,b},\rho\rangle \mu_{U,b},
\end{equation}
which matches the structure of the frame superoperator defined in~\cref{eq:frame_operator_def}.

\parTitle{Equivalence: step 3}
From the considerations above, it is now clear that the classical shadows, which read in terms of the POVM $(\mu_{U,b})$ as $\hat\rho=\mathcal M^{-1}(\mu_{U,b})$, are the elements of the canonical dual frame of the measurement frame.
This shows that the formalism to compute classical shadows with random unitary rotations and projective measurement follows as a special case of the general procedure for measurement frames outlined in~\cref{sec:shadow_tomography_formalism}.

\parTitle{Equivalence of the formalisms}
At first glance this procedure might still appear different from the one discussed in~\cref{sec:shadow_tomography_formalism}, as we did not explicitly use rescaled measurement frames here.
This is due to the covariant measurements being such that $\trace(\mu_{U,b})=1$ for all $U,b$, making the rescaling factors used in the definition $\tilde\calF_{I/d}$ unnecessary in these cases.
It follows that $\calM$ and $\tilde\calF_{I/d}$ only differ by the proportionality constant $d$.
These observations show that the formalism of shadow tomography via random unitary rotations can be seen as a direct application of the general formalism we present to rank-1 POVMs of the form $\mu_\psi=p_\psi\PP_\psi$ for some distribution over the states $\ket\psi$.
A direct numerical comparison between the results of applying our formalism to estimate observables from MUB measurements and the approach with uniformly random unitaries is presented in~\cref{fig:estimator_values_probabilities}.
As clearly shown in the figures, while the distribution of the estimators differs considerably in the two cases, the induced sample means have similar distributions, and both converge to the same Gaussian in the limit of infinite statistics.

\parTitle{Variance and shadow norm}
In Ref.~\cite{huang2020predicting}, the variance of the estimators for observables is bounded in terms of their so-called ``shadow norm'', which is there defined as
\begin{equation}\label{eq:def_shadow_norm}
\begin{aligned}
    \|\calO\|_{\rm sh}&\equiv \max_\sigma \bigg( \mathbb{E}_{U\sim\mathcal U}\sum_b
    \langle b|U\sigma U^\dagger|b\rangle\\
    &\times    \langle b|U \calM^{-1}(\calO) U^\dagger|b\rangle^2
    \bigg)^{1/2},
\end{aligned}\end{equation}
where the maximization is performed with respect to all possible states $\sigma$.
This expression is equivalent to
\begin{equation}\label{eq:shadow_norm_general_expr}
    \|\calO\|^2_{\rm sh} = \max_\sigma
    \sum_b \langle\mu_b,\sigma\rangle \langle\calO, \tilde\mu_b^{\rm can}\rangle^2.
\end{equation}
in the special case of $\bs\mu$ being the covariant measurement, \textit{i.e.} mapping $b\to (U,b)$ and $\mu_b\to \mu_{U,b}\equiv U^\dagger \mathbb{P}_b U$, and with $\tilde\mu_b^{\rm can}$ the canonical estimator associated to this measurement as given in~\cref{eq:definition_canonical_estimator}, which in this case reads $\tilde\mu_{U,b}^{\rm can}=d \calF_{I/d}^{-1}(\mu_{U,b})$.
Note that the explicit expression for $\calF_{I/d}$ for this POVM is
\begin{equation}
    \calF_{I/d} = d \int_{\mathbf U(d)} dU \sum_b \mathbb{P}(\mu_{U,b}),
\end{equation}
where we used $\trace(\mu_{U,b})=1$ for all $U,b$.
Therefore in terms of the operator $\calM$ defined in~\cref{eq:definition_M_op} we have $\calF_{I/d} = d \calM$, and $\tilde\mu_{U,b}^{\rm can} = \calM^{-1}(\mu_{U,b})$.
We finally recover~\cref{eq:def_shadow_norm} observing that $\calM$, or equivalently $\calF_{I/d}$, is Hermitian as a superoperator, and thus
\begin{equation}\small
    \langle b| U \calM^{-1}(\calO)U^\dagger |b\rangle =
    \trace(\mu_{U,b} \calM^{-1}(\calO))
    = \trace(\calO \tilde\mu_{U,b}^{\rm can}).
\end{equation}
Rewriting the shadow norm as in~\cref{eq:shadow_norm_general_expr} clearly shows that it corresponds to the nontrivial part of the variance, maximized over the input states, and that the definition of shadow norm is thus applicable for any choice of measurement and estimator.
In fact, we have the general result $\|\calO\|_{\rm sh}^2=\|A\|_{\rm op}$ with $A$ the operator defined in~\cref{eq:definition_A_op}, and thus the scaling results derived for random (Haar or Clifford) unitaries can be viewed as a particular instance of the more general results presented in~\cref{sec:bounds_averaged_variance}.

\begin{table*}[tbp]
    \centering
    \setlength{\tabcolsep}{7pt}
    \hypersetup{
    linkcolor=MidnightBlue,
    filecolor=magenta,      
    urlcolor=cyan
    }
    \begin{tabular}{clccc}
        \toprule
         & & Frame operator & Estimator & Variance of estimator
        \\ \midrule
        \multirow{2}{*}[-1ex]{State estimation} &
        (with prior $\rho$) &
        $\calF_{\rho} \equiv \sum_b \frac{\mathbb{P}(\mu_b)}{\langle\mu_b,\rho\rangle}$ &
        $\tilde\mu_b^{{(\rho)}} \equiv \frac{\calF_{\rho}^{-1}(\mu_b)}{\langle\mu_b,\rho\rangle}$ 
        &
        $\calE_\rho = \trace (\calF_{\rho}^{-1})-\trace(\rho^2)$  \\ \noalign{\smallskip}
        & (average over $\rho$) & 
        $\calF_{I/d}\equiv d\sum_b \frac{\mathbb{P}(\mu_b)}{\trace(\mu_b)}$ &
        $\tilde\mu_b^{\rm can} \equiv \frac{ d\calF_{I/d}^{-1}(\mu_b)}{ \trace(\mu_b) }$  &
        $\overline{\calE}_{P} = \trace(\calF_{I/d}^{-1})-P$  \\ \noalign{\smallskip}
        \hline \noalign{\smallskip}
        \multirow{3}{*}[-1ex]{\parbox{2cm}{Observable estimation}} & (with prior $\rho$) & $\calF_{\rho}$ & $\langle\calO,\tilde\mu_b\rangle
        = \langle \calO, \tilde\mu_b^{(\rho)}\rangle$  & $\on{Var}[\hat o|\rho] =\langle\calO,\calC_{\rho}(\calO)\rangle$ \vspace{1.5pt}\\ \noalign{\smallskip}
        & (average over $\rho$) & $\displaystyle\calF_{I/d}$ & $\langle\calO,\tilde\mu_b\rangle
        = \langle \calO, \tilde\mu_b^{\rm can}\rangle$  & $\overline{\on{Var}[\hat o|P]} =\langle\calO,\overline{\calC}_{\rho}(\calO)\rangle$ \\ \noalign{\smallskip}
        & (average over $\rho$, $\calO$) & $\calF_{I/d}$ & $\tilde\mu^{\rm can}$ & $\overline{\overline{\on{Var}[\hat o|P]}} = \frac{Vd}{d^2-1}\left[ \trace(\calF^{-1}_{I/d})-P\right]$ 
        \\ \noalign{\smallskip}
        \specialrule{.14em}{.1em}{.1em} 
        \noalign{\smallskip}
        \multicolumn{2}{c}{\multirow{2}{*}[-0.5ex]{\parbox{3cm}{Tight rank-1 POVM $\mu_b=w_b \mathbb{P}(\psi_b)$}}}  & \multirow{2}{*}{$ \calF_{I/d} = d\frac{\mathbb{P}(I) + \on{Id} }{d+1}$} & \multirow{2}{*}{$\tilde\mu_b^{\rm can} = (d+1) \mathbb{P}(\psi_b) - I$} & $\overline{\calE}_P = d^2+d-1-P$ \\
        & & & &$ \overline{\on{Var}[\hat o|P]} = \frac{Vd \left(d^2+d-1-P\right)}{d^2-1}$ 
        \\ \bottomrule
    \end{tabular}
    \caption{\textbf{Summarizing of the introduced quantities.}
    A schematic review of the expressions provided in the text for frame operator ($\calF$), state estimator ($\tilde{\mu}$), and associated variances.
    The first two rows summarize some of the quantities associated with the tomographic estimation of input states.
    Similarly, the next three rows refer to the case of recovering the expectation value of some target observable $\calO$.
    First and third rows summarize the quantities associated with the estimators that have minimum variance when the true input state is $\rho$.
    Second and fourth rows summarize the quantities associated with estimators that have minimum variance on average over the possible input states --- or equivalently, that have minimum variance when the true input state is $I/d$.
    The fifth row contains quantities associated with the estimator with minimum variance on average over both input states and target observables.
    Finally, the last row gives the explicit expressions for canonical frame operator, canonical estimator, MSE matrix, and averaged variance, in the special case of tight rank-1 measurement frames.
    }
    \label{table:summary_of_eqs}
\end{table*}

\section{Conclusions and forward look}
\label{sec:conclusions}

We have demonstrated how the general theory of measurement frames embodies a natural framework for shadow tomography. In doing so, we have assessed thoroughly the interplay between general measurements and associated optimal estimators to recover expectation values of target observables. Our results push the current knowledge in this context, recovering previously reported seminal results (cf. Ref.~\cite{huang2020predicting}) as special cases of our general framework, providing a natural understanding of the notion of shadow norm often used in the topical literature, and allowing estimation of finite sets of rank-1 bounded observables with a number of samples that does not grow with the dimension of the underlying space.
We provided analytical bounds for the estimation variance in several cases of interest, including the variance averaged over input states, and the variance averaged over both input states and target observables.
Among other things, we provided explicit results for the averaged variance in the case of tight measurement frames, general bounds tying the average variance to how close a POVM is to being a tight measurement frame, and also found an explicit expression for the non-averaged variance for rank-1 POVMs that form 3-designs.
In Table~\ref{table:summary_of_eqs} we provide a useful summary of some of the main expressions for frame operators and variances discussed throughout the paper.
To further ease the understanding of the different notions introduced in the manuscript, we also included in~\cref{app:toy_examples} several toy examples in which we explicitly work out frame superoperators and other relevant quantities.

Besides improving our understanding of general shadow tomography protocols, our results help the analysis and assessment of estimation errors in general measurement protocols, providing a unifying framework to understand both linear state tomography and shadow tomography. Our work thus contributes to the design of optimal strategies for single-setting quantum state tomography, which has recently attracted significant attention~\cite{stricker2022experimental,garcia2021learning,guerini2021quasiprobabilistic,garcia2022virtual}%
, as well as more general experimental protocols relying on learning properties of input states from measurement outcomes~\cite{suprano2021dynamical,suprano2021enhanced,zia2023regression,suprano2023experimental,tran2022measuring}%
.
Another context where our results will prove useful is the analysis of quantum reservoir computing architectures, which have been recently shown to be representable via generalized measurements summarizing the properties of the reservoir, and to be applicable for quantum state estimation tasks~\cite{innocenti2023potential}.
More generally, our formalism can be applied to any scenario where the goal is to extract properties of states from measurement outcomes, especially (although not exclusively) when the goal is to efficiently extract few properties from high-dimensional states.
Other potential avenues for research in this context include a more thorough exploration of the performance guarantees for 2-designs that are not also 3-designs, which would significantly expand the class of experimental situations where efficient dimension-independent estimation is possible.

By demonstrating the connection between the computation of classical shadows and the associated unbiased linear estimators, our approach establishes useful connections with metrology and estimation theory.
In particular, estimating only certain properties of an unknown quantum state is formally a quantum semiparametric estimation problem~\cite{Tsang2019} --- also known in the finite-dimensional case as estimation with nuisance parameters~\cite{Suzuki2019a}.
While quantum estimation is most commonly studied in a local/asymptotic scenario, we hope our approach will lead to further connections between shadow tomography and semiparametric estimation in the non-asymptotic regime.
Another intriguing area where an approach based on infinite-dimensional measurement frames could provide useful insights is continuous variable shadow tomography, which has only very recently been proposed~\cite{Gandhari2022,Becker2022a}.
Finally, the agility of the framework we put forward holds the premises to inform experimental efforts aimed at demonstrating a resource-inexpensive route to quantum state and property reconstruction.

\acknowledgments

LI acknowledges support from MUR and AWS under project 
PON Ricerca e Innovazione 2014-2020, ``Calcolo quantistico in dispositivi quantistici rumorosi nel regime di scala intermedia" (NISQ - Noisy, Intermediate-Scale Quantum).
IP is grateful to the MSCA COFUND project CITI-GENS (Grant nr. 945231).
FA acknowledges financial support from MUR under the PON Ricerca e Innovazione 2014-2020 project EEQU.
MP acknowledges the support by the European Union's Horizon 2020 FET-Open project  TEQ (Grant Agreement No.\,766900), the Horizon Europe EIC Pathfinder project  QuCoM (Grant Agreement No.\,101046973), the Leverhulme Trust Research Project Grant UltraQuTe (grant RPG-2018-266), the Royal Society Wolfson Fellowship (RSWF/R3/183013), the UK EPSRC (EP/T028424/1), and the Department for the Economy Northern Ireland under the US-Ireland R\&D Partnership Programme (USI 175 and USI 194).

\appendix

\newpage
\section{Properties of frame superoperators}
\label{sec:properties_frame_ops}

In this Section, we briefly review some important properties of the frame superoperators used in the paper.

\parTitle{Definition}
The frame superoperator that provides the minimum-variance state estimator when the true input is some reference state $\rho$ is
\begin{equation}
    \calF_{\rho} = \sum_b \frac{\mathbb{P}(\mu_b)}{\langle\mu_b,\rho\rangle},
\end{equation}
where we denote with $\mathbb{P}(\mu_b)$ the quantum map sending operators $X$ to $\mu_b \langle \mu_b,X\rangle$.
If $\rho\in\mathrm D(\mathbb{C}^d)$ is a $d$-dimensional state, then $\mu_b\in\on{Pos}(\mathbb{C}^d)$, and $\calF_{\rho}: \on{Lin}(\mathbb{C}^d)\to\on{Lin}(\mathbb{C}^d)$,
$\calF_{\rho}\in\on{Lin}(\on{Lin}(\mathbb{C}^d))$.
Being a linear function defined on linear operators, $\calF_{\rho}$ is a quantum map.
To connect to the more general theory of frames in linear algebra, this map is the frame operator corresponding to the rescaled frame of operators with elements $\{\mu_b/\sqrt{\langle\mu_b,\rho\rangle}\}_b$.

\parTitle{General properties}
Thinking of $\calF_{\rho}$ as a linear operator, we define its trace in the standard way, that is,
\begin{equation}
    \trace(\calF_{\rho}) = \sum_\alpha \langle \sigma_\alpha, \calF_{\rho}(\sigma_\alpha)\rangle,
\end{equation}
for an arbitrary orthonormal basis of Hermitian operators $\{\sigma_\alpha\}_{\alpha=1}^{d^2}$.
In particular, $\operatorname{tr}(\mathbb{P}(\mu_b))=\langle\mu_b,\mu_b\rangle=\trace(\mu_b^2)$, and thus
\begin{equation}\label{eq:trace_optimal_frame_op}
    \trace(\calF_{\rho}) =
    \sum_b \frac{\trace(\mu_b^2)}{\langle\mu_b,\rho\rangle}.
\end{equation}
We can furthermore verify by direct substitution that
\begin{equation}\label{eq:Frho_rho_equal_rho}
    \calF_{\rho}(\rho) = I,
    \qquad
    \calF_{\rho}^{-1}(I) = \rho.
\end{equation}

\parTitle{Properties of the inverse}
As discussed in the main text and derived in~\cref{sec:optimal_estimator_state}, the minimum-variance unbiased estimator provided by $\calF_{\rho}$ is $\hat f(b)\equiv\tilde\mu_b^{(\rho)}$ with
\begin{equation}\label{eq:optimal_dual_estimator_def}
    \tilde\mu_b^{(\rho)} \equiv \frac{1}{\langle\mu_b,\rho\rangle} \calF_{\rho}^{-1}(\mu_b).
\end{equation}
In particular, this means that the canonical dual frame corresponding to this frame operator has elements $\{\sqrt{\langle\mu_b,\rho\rangle} \tilde\mu_b^{(\rho)} \}_b$, and
\begin{equation}
    \calF_{\rho}^{-1} = \sum_b \mathbb{P}(\sqrt{\langle\mu_b,\rho\rangle } \tilde\mu_b^{(\rho)} )
    = \sum_b \langle\mu_b,\rho\rangle
    \mathbb{P}(\tilde\mu_b^{(\rho)}).
\end{equation}
Taking the trace, we obtain
\begin{equation}
    \trace(\calF_{\rho}^{-1}) = 
    \sum_b \langle\mu_b,\rho\rangle \trace( (\tilde\mu_b^{(\rho)})^2 ).
\end{equation}
This expression is particularly useful in that it directly enters the corresponding MSE matrix.

\parTitle{Canonical estimator}
The minimum-variance unbiased estimator when no prior knowledge about the true input state is assumed is obtained by setting $\rho=I/d$ in the frame superoperator.
We show in~\cref{app:sec:tightmeasurements_vs_weighted2designs} that the unbiased state estimator that minimizes the $L_2$ error averaged over unitarily equivalent states is $\hat f(b)\equiv \tilde\mu_b^{\rm can}$ with
\begin{equation}
    \tilde\mu_b^{\rm can} = \frac{d \calF_{I/d}^{-1}(\mu_b)}{\trace(\mu_b)}.
\end{equation}
The map $\calF_{I/d}$ has some further properties compared with its general counterpart. In particular, we have $\calF_{I/d}(I) = d I$, which means that $I$ is an eigenvector of $\calF_{I/d}$. This observation can be exploited to write the general decomposition
\begin{equation}\label{eq:decomposition_FId_for_general_measurements}
    \calF_{I/d} = d \mathbb{P}(I/\sqrt d) + \tilde\calF_{I/d},
\end{equation}
where $\tilde\calF_{I/d}$ is defined as the projection of $\calF_{I/d}$ on the subspace of traceless operators, that is,
\begin{equation}
    \tilde\calF_{I/d} =
    \Pi_{H_0} \calF_{I/d} \Pi_{H_0} =
    \Pi_{H_0} \tilde\calF_{I/d} \Pi_{H_0},
\end{equation}
where $\Pi_{H_0}\equiv \on{Id} - \mathbb{P}(I/\sqrt d)$ is the (superoperator) projector onto the subspace of traceless operators.
We employ the rescaled identity operator $I/\sqrt d$ in these expressions to ensure the normalization of the corresponding operator with respect to the Hilbert-Schmidt inner product: $\| I/\sqrt d\|_2\equiv \trace((I/\sqrt d)^2)=1$.
This decomposition also translates into corresponding simplified expressions for inverse and trace
\begin{equation}
\begin{aligned}
    \calF_{I/d}^{-1} &= \frac1d \mathbb{P}(I/\sqrt d) + \tilde\calF_{I/d}^{-1}, \\
    \trace(\calF_{I/d}^{-1}) &=
    \frac1d + \trace(\tilde\calF_{I/d}^{-1}).
\end{aligned}
\end{equation}
As discussed in more detail in~\cref{app:sec:tightmeasurements_vs_weighted2designs}, these expressions simplify even further in the special case of tight rank-1 measurement frames.

\parTitle{MSE matrix}
Following~\cite{zhu2014quantum}, we define the \textit{MSE matrix} corresponding to a state $\rho$, measurement $\mu$, and estimator $\tilde\mu$, as
\begin{equation}\label{eq:definition_MSEMatrix}
    \calC_\rho = \sum_b \langle\mu_b,\rho\rangle \mathbb{P}(\tilde\mu_b)
    - \mathbb{P}(\rho).
\end{equation}
Using the \textit{minimum-variance dual estimator} given in~\cref{eq:optimal_dual_estimator_def} the MSE matrix takes the simplified form
\begin{equation}
    \calC_\rho^{\rm opt} = \calF_\rho^{-1} - \mathbb{P}(\rho).
\end{equation}
For an arbitrary choice of possibly suboptimal estimator, we have the inequality $\calC_\rho \ge \calC_\rho^{\rm opt}$.
A remarkable property of the MSE matrix is that its trace equals the average $L_2$ state estimation error, as will be further discussed in the following Sections. The optimal MSE matrix can also be regarded as the (classical) Fisher information matrix, when the states are considered parametrized via their coefficients in some orthonormal basis.

\section{Minimum-variance state estimators}
\label{sec:optimal_estimator_state}
Let us consider a generic unbiased estimator --- or equivalently, as discussed before, a generic dual measurement frame --- and ask what is the associated average estimation error.
Measuring the error in the Hilbert-Schmidt distance we find
\begin{equation}\label{eq:variance_generic_dualframe}\small
\begin{gathered}
    \mathbb{E}[ \|\hat f-\rho\|_2^2] \equiv
    \sum_b \langle\mu_b,\rho\rangle
    \|\hat f(b)-\rho\|_2^2
    =
    \mathbb{E} \trace(\hat f^2) - \trace(\rho^2), \\
    \mathbb{E} \trace(\hat f^2) \equiv
    \Delta^2(\rho,\mu,\tilde\mu) \equiv \sum_b \langle\mu_b,\rho\rangle \trace(\tilde\mu_b^2),
\end{gathered}
\end{equation}
where we introduced the notation $\Delta^2\equiv\Delta^2(\rho,\mu,\tilde\mu)$ to denote the component of the average error that depends on the choice of measurement $\mu$ and dual $\tilde\mu$. The dependence of this quantity on these choices will not the explicitly shown in the following in order to ease the notation.

\parTitle{Minimum-variance dual frame}
As previously mentioned, different dual frames generally exist, and from~\cref{eq:variance_generic_dualframe} we can see that the choice of dual frame $\tilde\mu$ can affect the associated average estimation error.
It is then natural to ask what is the choice of dual frame that minimizes the estimation variance. This issue is addressed in~\cite{scott2006tight,zhu2011quantum,zhu2012phdthesis,zhu2014tomographic,perinotti2007optimal,dariano2007optimal}. We include here a different approach to deriving the minimum-variance unbiased estimators from the rescaled frame superoperator, using the method of Lagrange multipliers to directly perform the optimization with respect to all possible linear unbiased estimators.

\parTitle{Problem definition in vectorized notation}
To find the minimum-variance estimator $\tilde{\bs\mu}$, we observe that the task involves optimizing a quadratic function under linear constraints.
To see this more clearly, we temporarily neglect the fact that the various objects in~\cref{eq:variance_generic_dualframe} are operators, and simply think of them as vectors, upon some choice of orthonormal basis for the underlying Hilbert space.
The error term $\Delta^2$, which is what we need to minimize, can be written in vectorized notation as
\begin{equation}
    \sum_b \langle\mu_b,\rho\rangle \trace(\tilde\mu_b^2)
    = \sum_b \langle\mu_b,\rho\rangle \|\tilde\mu_b\|^2
    = \sum_{b,i,j} \mu_{bi} \rho_i \tilde\mu_{bj}^2,
\end{equation}
and the minimization must be performed with respect to the real parameters $\tilde\mu_{bj}$. More explicitly, this notation amounts to decomposing the operators as
\begin{equation}\label{eq:vectorized_opt_cost}
    \tilde\mu_{bj}\equiv \langle\sigma_j,\tilde\mu_b\rangle,
    \quad
    \mu_{bj}\equiv \langle\sigma_j,\mu_b\rangle,
    \quad
    \rho_i \equiv \langle\sigma_i,\rho\rangle,
\end{equation}
for some fixed choice of orthonormal operatorial basis $\{\sigma_i\}$.

We need to take into consideration that not all sets of parameters $\tilde\mu_{bj}$ correspond to a valid dual frame of $\mu$.
The definition of dual frame can be written in vectorized notation as
\begin{equation}
    \sum_{b,i} \mu_{bi} \rho_i \tilde\mu_{bj} = \rho_j,
\end{equation}
and this must hold for all possible choices of $\rho$.
Although these are in principle an infinite amount of constraints, they can be thought of as equivalent to the finite set of constraints corresponding to using as $\rho$ the elements of the considered operatorial basis $\{\sigma_i\}$. These constraints read
\begin{equation}\label{eq:vectorized_opt_linear_constraints}
    \sum_{b} \mu_{bi} \tilde\mu_{bj} = \delta_{ij}, \,\, \forall i,j.
\end{equation}
Let us denote this set of constraints as $\phi_{ij}\equiv \phi_{ij}(\mu,\tilde\mu)=0$, having defined
\begin{equation}
    \phi_{ij} \equiv \sum_b \mu_{bi}\tilde\mu_{bj} - \delta_{ij}.
\end{equation}

\parTitle{Lagrange multipliers to find stationary points}
To find the minimum of~\cref{eq:vectorized_opt_cost} under the constraints in~\cref{eq:vectorized_opt_linear_constraints}, we can use the general method of Lagrange multipliers.
For there to be a stationary point for the cost function under the given constraints, the gradient of the cost must be in the linear span of the gradients of the constraints. More explicitly, this means that there must be a set of coefficients $\lambda_{ij}$ such that, for all $b,k$, we have
\begin{equation}\label{eq:lagrange_mult_condition}
    \frac{\partial \Delta^2}{\partial\tilde\mu_{bk}}  =
    \sum_{ij} \lambda_{ij} 
    \frac{\partial \phi_{ij} }{\partial\tilde\mu_{bk}}.
\end{equation}
Computing the derivatives explicitly we find
\begin{equation}
\begin{gathered}
    \frac{\partial \Delta^2}{\partial\tilde\mu_{bk}}
    = 2\sum_i \mu_{bi} \rho_i \tilde\mu_{bk}, \\
    \frac{\partial \phi_{ij} }{\partial\tilde\mu_{bk}}
    = \mu_{bi} \delta_{jk},
\end{gathered}
\end{equation}
and thus~\cref{eq:lagrange_mult_condition} becomes
\begin{equation}\label{eq:vectorized_stationary_point_condition}
    2\sum_i \mu_{bi} \rho_i \tilde\mu_{bk} =
    \sum_i \lambda_{ik}  \mu_{bi}.
\end{equation}
Thinking of $\lambda,\mu,\tilde\mu$ as matrices, and defining the diagonal matrix $\Lambda$ with components $\Lambda_{ab}\equiv \delta_{ab} \langle\mu_b,\rho\rangle$,~\cref{eq:vectorized_stationary_point_condition,eq:vectorized_opt_linear_constraints} can be written concisely as
\begin{equation}\label{eq:concise_expr_lagrange_multipliers}
    2 \Lambda \tilde\mu = \mu \lambda,
    \qquad
    \mu^T \tilde\mu=I.
\end{equation}
Putting these together, and assuming $\Lambda$ to be invertible --- which amounts to using $\rho$ such that $\langle\mu_b,\rho\rangle>0$ for all $b$ --- we get
$
    2 I = 2 \mu^T \tilde \mu =
    \mu^T \Lambda^{-1} \mu \lambda
$. 
We thus conclude that the set of coefficients $\lambda_{ij}$ must have the form
\begin{equation}
    \lambda = 2 (\mu^T \Lambda^{-1} \mu)^{-1}.
\end{equation}
In writing this, we are interpreting $\lambda$ as a matrix, that is, as a linear operator in the underlying Hilbert space of Hermitian operators.
In other words, we can in this context interpret the set of Lagrange multipliers as a quantum map satisfying the given relations.
We can safely talk about the inverse of $\mu^T \Lambda^{-1}\mu$ because the corresponding map is invertible provided that $\mu$ is an IC-POVM.
This is because $\mu^T \Lambda^{-1} \mu$, going back to the original formalism in terms of operators, corresponds to the map
\begin{equation}\label{frame_operator_optimized_sum_projectors}
    \calF_\rho\equiv \sum_b \frac{ \mathbb{P}(\mu_b) }{ \langle\mu_b,\rho\rangle },
\end{equation}
and if $\{\mu_b\}$ is an IC-POVM then its elements span the space, and the quantum map thus defined is invertible.

With this solution for $\lambda$, we can now find the minimum-variance dual frame $\tilde\mu$ using~\cref{eq:concise_expr_lagrange_multipliers} as
\begin{equation}\label{eq:solution_optimal_dual_frame_matrix_notation}
    \tilde\mu = \Lambda^{-1}\mu (\mu^T \Lambda^{-1} \mu)^{-1}.
\end{equation}
Note that $\mu$ is not in general an invertible, nor squared, matrix, and thus we cannot simplify the inverse $(\mu^T \Lambda^{-1}\mu)^{-1}$ using the inverse of its elements.

Going back to the notation with operators, the minimum-variance dual frame we just found corresponds to the operators
\begin{equation}\label{eq:optimal_dual_frame}
\begin{gathered}
    \tilde\mu_b = \frac{1}{\langle\mu_b,\rho\rangle}
    \calF_\rho^{-1}(\mu_b),
\end{gathered}
\end{equation}
where we denoted with $\calF_\rho$ the map corresponding to the Lagrange multipliers, which can also be seen as the frame operator of the rescaled frame with elements $\mu_b/\sqrt{\langle\mu_b,\rho\rangle}$.
An explicit expression of $\calF_\rho^{-1}$ in terms of $\tilde \mu$ can be obtained using again~\cref{eq:concise_expr_lagrange_multipliers}: we get that $\lambda = 2\tilde{\mu}^T \Lambda \tilde{\mu}$ and therefore
\begin{equation}\label{frame_operator_optimized_inverse_sum_projectors}
    \calF_\rho^{-1} \equiv \sum_b \langle \mu_b, \rho \rangle\mathbb{P}(\tilde \mu_b),
\end{equation}
to be compared with $\calF_\rho$ of~\cref{frame_operator_optimized_sum_projectors}.

It is worth stressing the precise kind of ``optimality'' we just derived.
While the above optimal dual frame $\tilde\mu_b$ is an unbiased estimator with respect to all states, meaning $\sum_b \langle\mu_b,\rho\rangle\tilde\mu_b=\rho$ for all $\rho$, the associated estimation error and its optimality depend upon the specific state $\rho$ that is being examined.
Different choices of $\rho$ will correspond to different minimum-variance estimators, although all of these estimators are unbiased with respect to all states.
To find the estimator that has minimum-variance \textit{on average with respect to all possible input states} --- sampled uniformly from the Haar measure --- we just need to set $\rho = I/d$, obtaining
\begin{equation}
    \tilde\mu_b = \frac{d}{\trace(\mu_b)} \calF_{I/d}^{-1}(\mu_b),
\end{equation}
where
\begin{equation}\label{canonical_frame_operator_for_tomography}
\begin{gathered}
\calF_{I/d} \equiv d\sum_b \frac{\mathbb{P}(\mu_b)}{\trace(\mu_b)}, 
\quad 
\calF_{I/d}^{-1} \equiv \frac1{d} \sum_b \trace(\mu_b) 
\mathbb{P}(\tilde{\mu}_b).
\end{gathered}
\end{equation}
This can be deduced from the linearity of $\Delta^2$ in ~\cref{eq:variance_generic_dualframe} with respect to $\rho$. Therefore, integrating it over Haar-distributed states is equivalent to evaluating it at the maximally mixed state $\rho=I/d$.

\section{Minimum-variance observable estimators}
\label{sec:optimal_estimator_observables}

In~\cref{sec:optimal_estimator_state} we derived the form of the unbiased state estimator that minimizes the averaged $L_2$ estimation error.
The focus of shadow tomography protocols is, however, the estimation of observables, not retrieving tomographically complete descriptions of the state itself.
It would stand to reason that if the goal is estimating some target observable $\calO$, this might be possible with a different strategy that does not pass through state estimators, and gives even lower variance.
In this section, we will show that this is in fact not the case: any unbiased estimator $\hat o$ for an observable $\calO$, assuming it is unbiased for all possible input states, is bound to have the form $\hat o(b)=\langle \calO,\tilde\mu_b\rangle$ for some dual measurement frame $\tilde{\bs\mu}$.

\parTitle{All observable estimators pass through dual frames}
Let $\hat o$ be an unbiased estimator for a target observable $\calO$. This by definition means we have the relation
\begin{equation}
    \sum_b \hat o(b) \langle \mu_b,\rho\rangle = \langle\calO,\rho\rangle
\end{equation}
for all states $\rho$.
But by linearity of the inner product, this implies
    $\sum_b \hat o(b) \mu_b = \calO$,
which tells us that $\hat o(b)\in\mathbb{R}$ can be interpreted as the coefficients appearing in the expansion of $\calO$ as a linear combination of the frame elements $(\mu_b)_b$.
From the general theory of frames we then  conclude that there must be some dual frame $(\tilde\mu_b)_b$ such that
$\hat o(b)=\langle\tilde\mu_b,\calO\rangle$.
The opposite direction is immediate: if $\tilde\mu_b$ is a dual frame, and thus gives an unbiased state estimator, it is clear that $\langle\calO,\tilde\mu_b\rangle$ is an unbiased estimator for $\calO$.
We conclude that unbiased observable estimators always pass through some state estimator $\tilde{\bs\mu}$.

\parTitle{Minimum-variance observable estimators}
The above considerations tell us we can restrict our attention to estimators of the form $\hat o(b)=\langle\tilde\mu_b,\calO\rangle$.
The question remains as to what choice of estimator is best --- in the sense of having minimum variance --- to recover $\calO$ specifically.
To answer this question, we follow a reasoning similar to the one in~\cref{sec:optimal_estimator_state}.
If $\tilde\mu_b$ is a generic dual frame, with corresponding estimator $\hat o(b)\equiv\tilde\mu_b$, and $\rho$ is the true state, the variance reads
\begin{equation}
    \on{Var}[\hat o|\calO] =
    \sum_b \langle \mu_b,\rho\rangle \langle\calO,\tilde\mu_b\rangle^2 -
    \langle\calO,\rho\rangle^2.
\end{equation}
We focus on minimizing the first term with respect to $\tilde\mu$, as the second term only depends on $\rho$ and $\calO$.
In vectorized notation, the first term can be rewritten as
\begin{equation}~\label{eq:cost_optimization_observable}
    \sum_b \langle\mu_b,\rho\rangle \langle\calO,\tilde\mu_b\rangle^2 =
    \calO^T \tilde\mu^T \Lambda \tilde\mu \calO.
\end{equation}
In this notation $\tilde\mu$ and $\mu$ are matrices, $\Lambda$ is a diagonal matrix, and $\calO$ is a vector.
The constraints on the estimators remain $\tilde\mu^T \mu=\mu^T \tilde\mu = I$, which amounts to the set of constraints $\phi_{ij} = \sum_b \mu_{bi} \tilde\mu_{bj} - \delta_{ij}$.
Taking the derivative with respect to $\tilde\mu_{bk}$ on both cost function, given in~\cref{eq:cost_optimization_observable}, and constraints, we obtain that there must be coefficients $\lambda_{ij}$ such that
\begin{equation}
    2\sum_j \Lambda_{bb} \calO_k \tilde\mu_{bj} \calO_j =
    \sum_{ij} \lambda_{ij} \mu_{bi} \delta_{jk}.
\end{equation}
In more compact matrix notation, denoting with $\lambda$ the matrix with components $\lambda_{ij}$, we obtain the condition
\begin{equation}
    2\Lambda \tilde\mu\calO \calO^T=\mu\lambda.
\end{equation}
Multiplying both sides from the left first by $\Lambda^{-1}$ and then by $\mu^T$, and observing that $\mu^T\Lambda^{-1}\mu$ is the matrix representation of $\calF_{\rho}$, which is invertible for IC-POVMs, we find
\begin{equation}
    \lambda = 2(\mu^T\Lambda^{-1}\mu)^{-1} 
    \calO\calO^T.
\end{equation}
We thus conclude that the minimum-variance estimators are given by
\begin{equation}
    \tilde\mu \calO\calO^T =
    \Lambda^{-1}\mu(\mu^T \Lambda^{-1}\mu)^{-1} \calO\calO^T.
\end{equation}
More explicitly, this amounts to
\begin{equation}
    \sum_{k} \tilde\mu_{bk} \calO_k =
    \sum_{ik} \Lambda^{-1}_{bb}
    \mu_{bi} (\calF_\rho^{-1})_{ik} \calO_k.
\end{equation}
In operator notation, this reads
\begin{equation}
    \langle\calO, \tilde\mu_b\rangle =
    \frac{ \langle\calO,\calF_\rho^{-1}(\mu_b)\rangle }{\langle\mu_b,\rho\rangle}.
\end{equation}
We conclude that the estimators that minimize $\on{Var}[\hat o|\calO]$ when the input state is $\rho$, are all and only those such that
\begin{equation}
    \langle\calO,\tilde\mu_b\rangle
    = \langle \calO, \tilde\mu_b^{(\rho)}\rangle.
\end{equation}
In other words, the estimators equal to the minimum-variance state estimator $\tilde\mu_b^{(\rho)}$ on the span of $\calO$.
The associated variance can be written in terms of the MSE matrix as
\begin{equation}\small
    \langle \mathbb{P}(\calO),\mathcal C_\rho\rangle 
    \equiv
    \langle \calO, \calC_\rho(\calO)\rangle
    = 
    \sum_b \langle\mu_b,\rho\rangle \langle\mathcal O,\tilde\mu_b\rangle^2 -
    \langle\calO,\rho\rangle^2,
\end{equation}
where $\mathbb{P}(\calO)$ denotes the map $X\mapsto \langle\calO,X\rangle \calO$ for all $X\in\on{Lin}(\mathbb{C}^d)$.
We thus conclude that finding the state estimator giving an observable estimator with the smallest variance amounts to finding an estimator which acts like the overall minimum-variance state estimator on the support of the observable.
In other words, the minimum-variance state estimator also provides the minimum-variance observable estimator for any observable (under the same assumptions on the input state).
As in~\cref{sec:optimal_estimator_state}, all these result also hold in the averaged scenario: the estimators minimizing the variance on average over input states are obtained with the choice $\rho=I/d$, that is, using $\tilde{\bs\mu}^{\rm can}$.


\section{Tight measurements and weighted 2-designs}
\label{app:sec:tightmeasurements_vs_weighted2designs}

In this Section, we prove the equivalence between weighted complex projective 2-designs and tight measurement frames, discuss the general property of tight measurement frames, and prove the known lower bounds on $L_2$ average estimation error corresponding to canonical state estimators. Although using a slightly different formalism, the idea behind the proof reported here is analogous to the one reported in~\cite{roy2007weighted}.

\parTitle{Weighted 2-designs and tight measurement frames}
Consider a rank-1 measurement with elements $\mu_b=w_b \mathbb{P}(\psi_b)$, $b=1,...,m$, for some set of weights $w_b\in\mathbb{R}$ such that $\sum_b w_b=d$, and some set of vectors $\ket{\psi_b}\in\mathbb{C}^d$.
The corresponding canonical frame superoperator is by definition equal to
\begin{equation}
    \calF_{I/d} = d\sum_b \frac {\mathbb{P}(\mu_b) }{\trace(\mu_b)} =
    d\sum_b w_b \mathbb{P}(\mathbb{P}(\psi_b)),
\end{equation}
where we used $\trace(\mu_b)=w_b$, and we denoted with $\mathbb{P}(\mathbb{P}(\psi_b))$ the projector onto the projector $\mathbb{P}(\psi_b)$.
Here $\psi_b\in\mathbb{C}^d$ is a vector, $\mathbb{P}(\psi_b)\equiv \ketbra{\psi_b}{\psi_b}\in\on{Herm}(\mathbb{C}^d)$ is a linear operator projecting onto $\ket{\psi_b}$, and thus $\mathbb{P}(\mathbb{P}(\psi_b))$ is a linear operator acting in the space of linear operators, which projects onto the linear operator $\mathbb{P}(\psi_b)$. This object is a quantum map, which acts on any $X\in\on{Lin}(\mathbb{C}^d)$ as follows
\begin{equation}\small
    \mathbb{P}(\mathbb{P}(\psi_b)) (X) =
    \mathbb{P}(\psi_b) \langle \mathbb{P}(\psi_b),X\rangle \equiv
    \mathbb{P}(\psi_b) \langle \psi_b, X \psi_b\rangle.
\end{equation}
Being this a quantum map, we can consider its Choi representation. Given any map $\Phi:\on{Lin}(\calH_A)\to\on{Lin}(\calH_B)$, we define its Choi representation as the operator $J(\Phi)\in\on{Lin}(\calH_B\otimes\calH_A)$ such that
\begin{equation}
    J(\Phi) = \sum_{ij} \Phi(\ketbra{i}{j})\otimes \ketbra{i}{j}.
\end{equation}
For an arbitrary map of the form $\Phi(X)=\langle A,X\rangle B$ the Choi is $J(\Phi)=B\otimes \bar A$.
It follows that
\begin{equation}
    J( \mathbb{P}(\mathbb{P}(\psi_b)) ) =
    \mathbb{P}(\psi_b) \otimes \mathbb{P}(\psi_b)^T,
\end{equation}
and thus for the frame superoperator,
\begin{equation}
    J( \calF_{I/d} ) =
    d \left[ \sum_b w_b \mathbb{P}(\psi_b)^{\otimes 2} \right]^{T_B},
\end{equation}
where $T_B$ denoted the partial transpose of the second space.
This expression is useful because it provides a direct connection with the defining property of weighted 2-designs.
The vectors $\ket{\psi_b}$ form a complex projective 2-design with weights $w_b$ iff we have
\begin{equation}
    \sum_b w_b \mathbb{P}(\psi_b)^{\otimes 2} =
    d\frac{\Pi_{\rm sym} }{\binom{d+1}{2}}.
\end{equation}
The $d$ normalization factor on the right-hand side of this equation comes from $\sum_b w_b=d$, whereas in the standard definition of weighted 2-designs the weights are normalized to 1.
Using this relation we get
\begin{equation}
    J(\calF_{I/d})^{T_B} =
    d^2 \frac{\Pi_{\rm sym} }{\binom{d+1}{2} } =
    d\frac{ I\otimes I + W}{d+1},
\end{equation}
where we expressed the projector in terms of the Swap operator $W$ via $\Pi_{\rm sym}=(I+W)/2$.
Observing that $W^{T_B}= \sum_{ij} \ketbra{ii}{jj}$, $J(\mathbb{P}(I))=I\otimes I$, and $J(\on{Id})=W^{T_B}$, together with the fact that the Choi is a linear isomorphism between maps and operators, we conclude that
\begin{equation}\label{app:eq:frame_superop_tights}
    \calF_{I/d} = d\frac{\mathbb{P}(I) + \on{Id} }{d+1}.
\end{equation}
This derivation shows that, for any rank-1 IC-POVM with elements $\mu_b=w_b\mathbb{P}(\psi_b)$, the frame superoperator $\calF_{I/d}$ has this form \textit{if and only if} the vectors $\ket{\psi_b}$ and weights $w_b$ form a weighted 2-design.
\Cref{app:eq:frame_superop_tights} differs by a factor of $d$ to the expressions for tight frames found e.g. in~\cite{scott2006tight}, but that is simply due to the definitions of frame superoperator differing by a $d$ factor, and will not affect our results.

\parTitle{Properties of tight frame superoperators}
Suppose now $\mu$ is a tight rank-1 IC-POVM, and thus the frame superoperator satisfies~\cref{app:eq:frame_superop_tights}. In light of the decomposition of~\cref{eq:decomposition_FId_for_general_measurements}, we can rewrite the frame operator as
\begin{equation}
    \calF_{I/d} = d \mathbb{P}(I/\sqrt d) +\frac{d}{d+1} \left(\on{Id} -
    \mathbb{P}(I/\sqrt d)
    \right).
\end{equation}
This writing is useful because it splits the action of $\calF_{I/d}$ into two invariant orthogonal subspaces. The superoperators $\mathbb{P}(I)/d$ and $\on{Id}-\mathbb{P}(I)/d$ project onto the one-dimensional subspace spanned by $I$, and the $(d^2-1)$-dimensional subspace of traceless Hermitian matrices, respectively.
It follows that the inverse has the form
\begin{equation}\label{eq:FIdinv_tightframes}
\begin{aligned}
    \calF_{I/d}^{-1}& = \frac1 d \mathbb{P}(I/\sqrt d) + 
    \frac{d+1}{d} \left(\on{Id} - \mathbb{P}(I/\sqrt d)\right) \\
   & = \frac{(d+1)\on{Id} - \mathbb{P}(I) }{d}.
\end{aligned}
\end{equation}
Using $\tilde\calF_{I/d}$, defined as in~\cref{eq:decomposition_FId_for_general_measurements}, we then also obtain for tight measurement frames the expression:
\begin{equation}\label{eq:trace_tildeFcan_tightmeasurements}
    \trace(\tilde\calF_{I/d})
    = \frac{d(d^2-1)}{d+1} = d(d-1).
\end{equation}

\parTitle{Estimators for tight measurement frames}
Knowing the general structure of the optimal frame corresponding to a tight measurement with elements $\mu_b=w_b \mathbb{P}(\psi_b)$, we can compute explicitly the structure of the corresponding estimator $\hat f(b)\equiv \tilde\mu_b^{\rm can}$, which gives
\begin{equation}
    \tilde\mu_b^{\rm can} = \frac{d}{\trace(\mu_b)} \calF_{I/d}^{-1}(\mu_b) =
    (d+1) \mathbb{P}(\psi_b) - I.
\end{equation}

\parTitle{Lower error bounds for tight measurement frames}
We will show here that the $L_2$ estimation error averaged over unitarily invariant input states, when using any unbiased estimator, is lower bounded by $d^2+d-1-\trace(\rho^2)$, with the inequality saturated for rank-1 tight measurements. This was first proven in~\cite{scott2006tight,zhu2011quantum}.
To estimate the average state estimation errors we use the MSE matrix $\calC_\rho$ discussed in~\cref{eq:definition_MSEMatrix}.
If we assume the estimators $\tilde\mu_b$ do not depend on the input state --- as is the case for the canonical estimator, but not for the optimal ones --- then taking the uniform average with respect to states unitarily equivalent to $\rho$ we get
\begin{equation}
\begin{aligned}
    \overline{\calC}_{\rho} &=
    \sum_b \langle \mu_b,I/d\rangle \mathbb{P}(\tilde\mu_b) - \int_{\mathbf U(d)} dU\, \mathbb{P}(U\rho U^\dagger) \\
   & = \calF_{I/d}^{-1} -  \int_{\mathbf U(d)}dU\, \mathbb{P}(U\rho U^\dagger),
\end{aligned}
\end{equation}
where the integral is taken with respect to the uniform Haar measure in the group of unitary matrices.
Taking the trace we get the average error as
\begin{equation}
\begin{gathered}
    \overline\calE_\rho
    = \trace(\overline{\calC}_{\rho} )
    = \trace(\calF_{I/d}^{-1}) - \trace(\rho^2).
\end{gathered}
\end{equation}
For tight rank-1 measurement frames we know from~\cref{eq:FIdinv_tightframes} that
\begin{equation}
    \trace(\calF^{-1}_{I/d}) =d^2 + d -1.
\end{equation}
Let us now show that this is also the lower bound for an arbitrary measurement.
From~\cref{eq:trace_optimal_frame_op} we see that for any $\bs\mu$,
\begin{equation}
    \trace(\calF_{I/d})
    =
    d\sum_b \frac{\trace(\mu_b^2)}{\trace(\mu_b)}
    \le d\sum_b \trace(\mu_b) = d^2,
\end{equation}
where we used the inequality $\trace(X^2)\le \trace(X)^2$ for $X\ge0$, which is saturated iff $\on{rank}(X)=1$. Thus $\trace(\calF_{I/d})\le d^2$ and $\trace(\tilde\calF_{I/d})= \trace(\calF_{I/d})-d \le d(d-1)$ with equality for rank-1 measurements. But also, being $\tilde\calF_{I/d}$ Hermitian and non-singular as a linear (super)operator, we have
\begin{equation}
    \trace(\tilde\calF_{I/d}) = \sum_{k=1}^{d^2-1} \lambda_k,
    \qquad
    \trace(\tilde\calF_{I/d}^{-1}) = \sum_{k=1}^{d^2-1} \frac{1}{\lambda_k},
\end{equation}
where $\lambda_k$ are the eigenvalues of $\tilde\calF_{I/d}$, and there are $d^2-1$ terms in the sum because $\on{rank}(\tilde\calF_{I/d})=d^2-1$.
A direct application of Lagrange's multipliers then allows us to find the minimum value of $\trace(\tilde\calF_{I/d}^{-1})$ under the constraint of $\lambda_k\ge0$ and $\trace(\tilde\calF_{I/d})=d(d-1)$, which reads
\begin{equation}
\label{eq:trace_Ftilde_inverse_lower_bound}
    \trace(\tilde\calF_{I/d}^{-1}) \ge \frac{(d^2-1)(d+1)}{d},
\end{equation}
with equality holding iff all the eigenvalues have the same value, that is, iff $\tilde\calF_{I/d}$ is a multiple of the identity (when acting on the $(d^2-1)$-dimensional subspace of traceless Hermitian matrices).
We conclude that for \textit{any} measurement, we have the lower bound
\begin{equation} \label{eq:trace_F_tildeF}
    \trace(\calF_{I/d}^{-1}) = \frac1d+ \trace(\tilde\calF_{I/d}^{-1})
    \ge d^2+d-1,
\end{equation}
with the inequality saturated for tight rank-1 measurements.
We therefore just proved that for any measurement, the average $L_2$ estimation error when using the canonical estimator is lower bounded as
\begin{equation}
    \overline\calE_\rho \ge d^2+d-1 - \trace(\rho^2).
\end{equation}
It is also possible to study the errors corresponding to more general not-necessarily-rank-1 tight IC POVMs. This analysis can be found in~\cite{zhu2014tomographic}, and the smallest possible average $L_2$ estimation error, when the POVM elements have average purity $\wp$, works out to be
\begin{equation}
    \overline\calE_\rho = \frac{(d^2-1)^2}{d^2\wp - d}
    - \left[ \trace(\rho^2) - \frac1d \right],
\end{equation}
where
\begin{equation}
    \wp \equiv \frac1d\sum_b \frac{\trace(\mu_b^2)}{\trace(\mu_b)} = \frac{\trace(\calF_{I/d})}{d^2} \in [1/d,1].
\end{equation}

\section{Errors to estimate single observables}
\label{sec:errors:singleobservables}

As discussed in~\cref{app:sec:tightmeasurements_vs_weighted2designs}, to study the estimation errors associated to a given state estimator, it is useful to introduce the MSE matrix $\calC_\rho$. 
Suppose now we want to estimate the expectation value of some observable $\calO$ on a state $\rho$, using the unbiased estimator $\hat o(b)\equiv \langle\calO,\hat f(b)\rangle=\langle\calO,\tilde\mu_b\rangle$.
The associated variance is
\begin{equation}
    \on{Var}[\hat o|\rho,\calO,\bs\mu,\tilde{\bs\mu}] = \sum_b \langle\mu_b,\rho\rangle \langle\calO,\tilde\mu_b\rangle^2 - \langle\calO,\rho\rangle^2.
\end{equation}
As in the main text, the functional dependence on $\calO$, $\bs\mu$, and $\tilde{\bs\mu}$ will be left implicit for notational conciseness.
This variance can be expressed via the MSE matrix as
\begin{equation}
    \on{Var}[\hat o| \rho] = \langle \mathbb{P}(\calO), \calC_\rho\rangle \equiv \langle\calO,\calC_\rho(\calO)\rangle.
\end{equation}

\parTitle{Expression for averaged variance}
Let us focus on the behaviour of the variance when using the canonical state-independent estimator $\tilde\mu_b^{\rm can}$. With this choice, taking the average over input states with purity $P\equiv\trace(\rho^2)$, we have
\begin{equation}\label{eq:averaged_variance}
\begin{gathered}
    \overline{\on{Var}[\hat o|P]} \equiv
    \int_{\mathbf U(d)} dU\, \on{Var}[\hat o|U\rho U^\dagger,\calO] \\
    = \sum_b \langle \mu_b,I/d\rangle \langle\calO,\tilde\mu_b^{\rm can}\rangle^2 - \int_{\mathbf U(d)}dU \langle\calO,U\rho U^\dagger\rangle^2 \\
    = \langle\calO,\calF_{I/d}^{-1}(\calO)\rangle
    - \beta.
\end{gathered}
\end{equation}
where $\beta$ is the expectation value of $\langle\calO,\rho\rangle^2$ over states with purity $P$.
This quantity is computed using the known formulas to integrate polynomials in the components of unitaries matrices over the uniform Haar measure~\cite{collins2003moments}, and equals
\begin{equation}\label{eq:definition_of_beta2}
    \beta = \frac{\trace(\calO)^2}{d^2}
    + \frac{d P-1}{d^2-1} V,
\end{equation}
where $V\equiv \langle\calO^2\rangle-\langle\calO\rangle^2$ is the variance of the observable computed on the maximally mixed state, with $\langle \calO\rangle\equiv\trace(\calO)/d$ and $\langle\calO^2\rangle\equiv\trace(\calO^2)/d$.
Note that the averaged variance depends on $P$, but not on the specific choice of $\rho$.
Let us now focus on the term $\langle\calO,\calF_{I/d}^{-1}(\calO)\rangle$, which is the one depending on the POVM.
Using the decomposition in~\cref{eq:decomposition_FId_for_general_measurements} for $\calF_{I/d}$ we have
\begin{equation}\label{eq:decomposition_expval_Finv}
    \langle\calO,\calF_{I/d}^{-1}(\calO)\rangle =
    \frac{\trace(\calO)^2}{d^2} +
    \langle\calO,\tilde\calF_{I/d}^{-1}(\calO)\rangle.
\end{equation}
Putting together~\cref{eq:averaged_variance,eq:definition_of_beta2,eq:decomposition_expval_Finv}, we obtain the general expression for the averaged variance corresponding to the canonical estimator:
\begin{equation}\label{eq:averaged_variance_final_expr}
    \overline{\on{Var}[\hat o|P]} =
    \langle\calO,\tilde\calF_{I/d}^{-1}(\calO)\rangle -
    \frac{d P-1}{d^2-1} V.
\end{equation}

\parTitle{Bounds for the averaged variance}
The first term can be bounded in terms of the eigenvalues of $\tilde\calF_{I/d}^{-1}$, as
\begin{equation}\label{eq:error_observable_vs_minmaxeigvals}
    \frac{Vd}{ \lambda_{\rm max}(\tilde\calF_{I/d}) }
    \le 
    \langle\calO,\tilde\calF_{I/d}^{-1}(\calO)\rangle
    \le
    \frac{Vd}{ \lambda_{\rm min}(\tilde\calF_{I/d}) },
\end{equation}
where $\lambda_{\rm min}(\tilde\calF_{I/d}),\lambda_{\rm max}(\tilde\calF_{I/d})$ are smallest and largest eigenvalues of $\tilde\calF_{I/d}$ (which is positive definite as an operator whenever $\bs\mu$ is IC).
This bound is obtained observing that $\tilde{\calF}_{I/d}$, and therefore also $\tilde{\calF}_{I/d}^{-1}$, is a (Hermitian) linear operator acting on the space of $\on{Herm}(\mathbb{C}^d)$ spanned by traceless Hermitian operators.
In general, if $H\in\on{Lin}(V)$ is a Hermitian operator acting on some vector space $V$,
with support $W\equiv \on{supp}(H) \subseteq V$, 
then for any $v\in W$ we have
\begin{equation}
    \lambda_{\rm min}(H) \|v_W\|^2 \le 
    \langle v, Hv\rangle
    \le \lambda_{\rm max}(H) \|v_W\|^2,
\end{equation}
where $v_W$ is the projection of $v$ on $W$, and $\lambda_{\rm min}(H),\lambda_{\rm max}(H)$ are smallest and largest \textit{nonzero} eigenvalues of $H$.
Applying this with $H= \tilde{\calF}_{I/d}^{-1}$ and $v= \calO$ we get~\cref{eq:error_observable_vs_minmaxeigvals}, because the orthogonal projection of $\calO$ on the subspace of traceless Hermitian operators is $\calO-\trace(\calO)I/d$, and $\|\calO-\trace(\calO)I/d\|^2=Vd$.

\parTitle{General bounds for worst-case variance}
From~\cref{eq:error_observable_vs_minmaxeigvals} we get a general upper bound for the variance in the form:
\begin{equation}\label{eq:upperbound_averagevariance}
    \overline{\on{Var}[\hat o|P]}
    \le Vd\left[\frac{1}{\lambda_{\rm min}(\tilde\calF_{I/d})} - \frac{P-1/d}{d^2-1}\right].
\end{equation}
This upper bound still depends on $\calO$ via $V$, but this dependence is intrinsic to the observable --- it is the average variance one would obtain estimating $\langle \calO\rangle$ from projective measurements in its eigenbasis, and is thus the absolute lower bound achievable for $\overline{\on{Var}[\hat o|P]}$.
We can thus interpret~\cref{eq:upperbound_averagevariance} as the average variance corresponding to the hardest-to-estimate observable.
We will now attempt to provide more precise bounds for this quantity in terms of general symmetry properties of the POVM.
In particular, remembering that a POVM is tight \textit{iff} its frame superoperator satisfies $(d^2-1)\trace(\tilde\calF_{I/d}^2)=\trace(\tilde\calF_{I/d})^2$, a natural choice is to explore the set of IC-POVMs under the constraints $\trace(\tilde\calF_{I/d})=a$ and $\trace(\tilde\calF_{I/d}^2)=b$ for some given $a,b>0$.

We then analyze what is the smallest possible value of the average variance for the hardest-to-estimate observable, as a function of $a$ and $b$.
More formally, we therefore consider the following question: \textit{what is the POVM that gives the smallest $1/\lambda_{\rm min}(\tilde\calF_{I/d})$, under the above constraints?} This is equivalent to asking for the largest possible $\lambda_{\rm min}(\tilde\calF_{I/d})$ under the same constraints.
In turn, focusing on the eigenvalues, this question is equivalent to:
\textit{within the set of tuples $\lambda_1,...,\lambda_{d^2-1}>0$ such that $\sum_k\lambda_k=a$ and $\sum_k \lambda_k^2=b$, what is the largest possible value of $\min(\lambda_k)$?}
For consistency, the coefficients $a,b$ need to satisfy $0<b\le a^2\le b(d^2-1)$, which follows directly from the AM–GM inequality.

Solving this optimization problem is made somewhat more difficult by the cost function $\min(\lambda_1,...,\lambda_{d^2-1})$ being non-differentiable.
We can nonetheless convert it into a differentiable cost by introducing additional slack variables.
Let us for notational conciseness define $m\equiv d^2-1$. Our problem can be restated as that of maximizing $\lambda_1$, with respect to the $2m-1$ variables $\lambda_1, ...,\lambda_m,s_2,...,s_m$, subject to the constraints
\begin{equation}
\begin{gathered}
    \lambda_k\ge0, \quad
    \sum_{k=1}^m \lambda_k=a,
    \quad \sum_k\lambda_k^2=b, \\
    \lambda_1 + s_k^2 = \lambda_k, \quad\forall k=2,...,m.
\end{gathered}
\end{equation}
The constraints $\lambda_1+s_k^2=\lambda_k$ are introduced to enforce $\lambda_1\le \lambda_k$, and thus ensure that the solution to this problem corresponds to the solution of the original one.
Using the method of Lagrange multipliers~\cite{boyd2004convex}, define the Lagrangian function
\begin{equation}
\begin{gathered}
    L = \lambda_1 + \alpha \left(\sum_{k=1}^m\lambda_k-a\right) +\beta\left(\sum_{k=1}^m \lambda_k^2 - b\right) \\
    +
    \sum_{k=2}^m \gamma_k (\lambda_k-\lambda_1-s_k^2).
\end{gathered}
\end{equation}
Imposing $\nabla L=0$ gives the conditions
\begin{equation}\label{eq:conditions_alphabeta_lagrangemultipliers}
\begin{gathered}
    1 +  \alpha + 2\beta \lambda_1 - \sum_{k=2}^m \gamma_k=0,
    \\
    \alpha + 2\beta \lambda_k + \gamma_k =0,\quad\forall k\ge 2,
    \\
    \gamma_k s_k=0, \quad \forall k\ge2.
\end{gathered}
\end{equation}
We can explore the different sets of solutions compatible with these constraints by taking into account the number of coefficients $s_k$ that equal $0$:
\begin{enumerate}
    \item Suppose $s_2,\dots,s_m\neq0$. This implies $\gamma_2=\cdots=\gamma_m=0$, which in turns implies $\lambda_1 < \lambda_2$ and $\lambda_2=\cdots =\lambda_m$. The constraints in terms of $a,b$ simplify to $\lambda_1 + (m-1)\lambda_2=a$ and $\lambda_1^2+(m-1)\lambda_2^2=b$. These two equations give two solutions for $\lambda_1$, one of which is unfeasible because corresponds to $\lambda_1>\lambda_2$; the other one is feasible, and is a possible solution:
    \begin{equation}
        \lambda_1 = \frac{a}{m} - \frac{\sqrt{(m-1)(b m-a^2)}}{m}.
    \end{equation}
    \item More generally, suppose $s_2=\dots=s_\ell=0$ and $s_{\ell+1},\dots,s_m\neq0$ for some $2\le \ell\le m$.
    This implies $\gamma_{\ell+1}=\cdots = \gamma_m=0$, which in turn implies $\lambda_{\ell+1}=\cdots = \lambda_m$.
    Furthermore, $s_2=\cdots =s_\ell=0$ means that $\lambda_1=\dots = \lambda_\ell$. We therefore reduce again to a situation with only two distinct values for the coefficients $\lambda_k$, and the constraints again simplify to
    $\ell \lambda_1 + (m-\ell)\lambda_m=a$ and $\ell \lambda_1^2 + (m-\ell)\lambda_m^2=b$. Solving this and keeping the solution consistent with the constraints gives
    \begin{equation}\label{eq:general_sol_lambda1}
        \lambda_1 = \frac{a}{m} - \frac{\sqrt{\ell(m-\ell)(bm-a^2)}}{m \ell}.
    \end{equation}
\end{enumerate}
The above cover all possible scenarios, up to a permutation of the vanishing coefficients $s_k$ (any such permutation does not affect the resulting solution for $\lambda_1$ due to the problem symmetry).
The final solution is thus the maximum of~\cref{eq:general_sol_lambda1} for $\ell=1,...,m$. Observing that $\sqrt{\ell(m-\ell)}/\ell$ decreases monotonically with $\ell=1,...,m$, we conclude that the largest $\lambda_1$ is obtained when $\ell=m$. This case, however, corresponds to having $\lambda_1=\dots=\lambda_m=a/m$, which is only compatible with the constraints if $bm=a^2$.
The more general scenario is obtained for $\ell=m-1$, corresponding to having $\lambda_1=\dots=\lambda_{m-1}<\lambda_m$, and is possible for all $a,b>0$ with $a^2\le bm$.

To summarize, we concluded that the largest $\min(\lambda_1,...,\lambda_{m})$, $m\equiv d^2-1$, compatible with given values of $a=\trace(\tilde\calF_{I/d})$ and $b=\trace(\tilde\calF_{I/d}^2)$ is
\begin{equation}
    \lambda_1^* \equiv \frac{a}{m} - \frac{\sqrt{(m-1)(bm-a^2)}}{m(m-1)},
\end{equation}
which in the special case where $bm=a^2$, corresponding to $\tilde\calF_{I/d}$ being a multiple of the identity and thus $\bs\mu$ being a tight measurement frame, reduces to $\lambda_1^* = a/m$.
Reformulating this in terms of the variance, we concluded that, compatibly with $\bs\mu$ such that $\trace(\tilde\calF_{I/d})=a$ and $\trace(\tilde\calF_{I/d}^2)=b$, we have
\begin{equation}
    \max_{\cal O}
    \frac{\overline{\on{Var}[\hat o|P,\mathcal O]}}{Vd}
    \ge 
    \frac{1}{\lambda_1^*} - \frac{P-1/d}{d^2-1},
\end{equation}
with the inequality saturated by some POVM whose canonical estimator gives equal average variance for all observables (in some orthonormal basis of Hermitian operators) but one.
Furthermore, for tight measurements, $\tilde\calF_{I/d}$ is a multiple of the identity, $\trace(\tilde\calF_{I/d})=d(d-1)$ as per~\cref{eq:trace_tildeFcan_tightmeasurements}, $\lambda_1^*=\trace(\tilde\calF_{I/d})/(d^2-1)$, and thus
\begin{equation}\label{eq:app:conclusion_upperbound_obsestimationerror}
    \overline{\on{Var}[\hat o|P,\calO]}
    = Vd \left(\frac{d^2+d-1-P}{d^2-1}\right),
\end{equation}
where we made the dependence of $\calO$ explicit to point out that all observables give the same expression for the variance.
We recognize in particular the term $d^2+d-1-P$ which is the optimal state estimation $L_2$ error discussed in~\cref{app:sec:tightmeasurements_vs_weighted2designs}.
From~\cref{eq:app:conclusion_upperbound_obsestimationerror} we see that for tight rank-1 measurements, the asymptotic growth of the variance with the state dimension can be cancelled out by the choice of observable since it only depends on the factor $Vd$.
For example, for any observable that is a projection onto a pure state, $\calO=\mathbb{P}_\psi$ for some $\ket\psi$, we have $\trace(\calO^2)=\trace(\calO)=1$ , $Vd=(d-1)/d$, and therefore
\begin{equation}
    \overline{\on{Var}[\hat o|P,\mathbb{P}_\psi]}
    = \frac{d^2+d-1-P}{d(d+1)},
\end{equation}
where we now included the explicit dependence of the variance on the observable $\calO=\mathbb{P}_\psi$.
This gives $\overline{\on{Var}[\hat o|P,\mathbb{P}_\psi]}\to1$ for large $d$, regardless of $\ket\psi$, meaning the estimation errors to estimate such observables do not increase with the dimension of the space.
Similarly, for normalized observables $\calO_N$ with $\trace(\calO_N)=0$ and $\trace(\calO_N^2)=1$, we have $Vd=1$, and thus
\begin{equation}
    \overline{\on{Var}[\hat o|P,\calO_N]} =
    \frac{d^2+d-1-P}{d^2-1}.
\end{equation}
As a counterexample, if one studies the variance associated to estimating $\calO_n= \sigma_{i_1}\otimes \dots \otimes \sigma_{i_n}$ defined as a products of $n$ Pauli matrices, then $\trace(\calO_n)=0$, $\trace(\calO_n^2)=2^n$, $Vd=d=2^n$ and
\begin{equation}
    \overline{\on{Var}[\hat o|P,\calO_n]} =
    d\frac{d^2+d-1-P}{d^2-1} = O(d), \quad d\rightarrow \infty
\end{equation}
meaning the average variance increases linearly with $d$.

\section{Averaged error for the estimation of observables}
\label{sec:error_avg_observables}

We focus in this section on the variance averaged over both the input states at fixed purity, and over unitarily equivalent random target observables.

We already derived in~\cref{eq:averaged_variance} the expression for the variance averaged over unitarily equivalent input states, for any given fixed observable $\calO$.
Perform also an average over unitarily equivalent observables, we get
\begin{equation}\label{eq:average_variance_obs_state}
    \overline{\overline{\on{Var}[\hat o]}} \equiv \int_{\mathbf{U}(d)} dU \,
    \overline{\on{Var}[\hat o|U\calO U^\dagger]}
    \equiv \alpha' - \beta
\end{equation}
where
\begin{equation}\label{eq:average_variance_obs_state_alpha}\small
\begin{gathered}
    \alpha'=\frac{1}{d(d^2-1)} \sum_b \tr \mu_b \Bigg\{ \left( \trace \calO \right) ^2 \left[ \left(\trace \tilde{\mu}_b\right)^2- \frac{\trace \tilde{\mu}_b^2}{d} \right] + \\
    + \trace \calO^2 \left[ \trace \tilde{\mu}_b^2 - \frac{\left(\trace \tilde{\mu}_b\right)^2}{d} \right] \Bigg\} \\
    =
    \frac{\trace (\calF_{I/d}^{-1}) \left( d \trace \calO^2 - (\Tr \calO)^2 \right)
    + \left( d (\trace \calO)^2 - \trace \calO^2 \right)}{d(d^2-1)}
    \\
    = V \frac{d \trace(\calF_{I/d}^{-1})-1}{d^2-1}
    + \frac{(\trace \calO)^2}{d^2},
\end{gathered}
\end{equation}
and where $\beta$, given in~\cref{eq:definition_of_beta2}, does not change performing this second average since it only depends on $\calO$ via $\trace(\calO)$ and $\trace(\calO^2)$.
These expressions further simplify to
\begin{align}\label{eq:average_variance_state_obs_general}
    \overline{\overline{\on{Var}[\hat o|\calO]}} &=\frac{Vd}{d^2-1}\left( \trace(\calF^{-1}_{I/d})-P\right) \nonumber \\
    &=\frac{Vd}{d^2-1}\left(
     \trace(\tilde{\calF}_{I/d}^{-1})
    - P + \frac1d
    \right),
\end{align}
using the expression for $\beta$ given in~\cref{eq:definition_of_beta2}. It is instructive to compare this equation with the results of~\cref{sec:errors:singleobservables} and, for instance, with the upper bound of ~\cref{eq:upperbound_averagevariance}.
Using the lower bound on the trace given by~\cref{eq:trace_Ftilde_inverse_lower_bound}, we get
\begin{equation}\label{eq:average_variance_state_obs_lower_bound}
    \overline{\overline{\on{Var}[\hat o|\calO]}} \geq
    \frac{Vd \left(d^2+d-1-P\right)}{d^2-1},
\end{equation}
with equality \textit{iff} $\bs \mu$ is a tight rank-1 measurement.

\noindent\\
To provide some examples, let us consider observables of the form $\calO = \mathbb{P}_\psi$ for some $\ket{\psi}$, for which we have $Vd=(d-1)/d$ and thus, from~\cref{eq:average_variance_state_obs_general},
\begin{equation}
\label{eq:average_variance_state_obs_witness}
    \overline{\overline{\mathrm{Var}[\hat{o}|\mathbb{P}_{\psi}]}}
    = \frac1{d(d+1)}\left[ \trace \left( \tilde\calF_{I/d}^{-1} \right) - P + \frac1d \right].
\end{equation}
\Cref{eq:average_variance_state_obs_lower_bound} now reads
\begin{equation}
   \frac{2}{3} \leq \min_{\bs \mu} \bigg\{ \overline{\overline{\mathrm{Var}[\hat{o}|\mathbb{P}_{\psi}]}} \bigg\} = 1 - \frac{ 1+ P }{d(d+1)} \leq 1,
\end{equation}
which is an increasing function of $d$ but bounded from above, as expected for this type of observable.

\noindent\\
Similarly, for Pauli observables of the form $\calO_n = \sigma_{i_1}\otimes \dots \otimes \sigma_{i_n}$, acting on $n$ qubits ($d=2^n$), we have $Vd = d$ and therefore~\cref{eq:average_variance_state_obs_general} becomes
\begin{equation}\label{eq:average_variance_state_obs_pauli}
\small
    \overline{\overline{\mathrm{Var}[\hat{o}|\calO_n]}} = \frac{d}{d^2-1}
    \left[\trace \left( \tilde\calF_{I/d}^{-1} \right) - P + \frac1d \right].
\end{equation}
As before, this quantity is bounded from below by 
\begin{equation}
    \overline{\overline{\mathrm{Var}[\hat{o}|\calO_n]}} \geq  \min_{\bs \mu} \bigg\{ \overline{\overline{\mathrm{Var}[\hat{o}|\calO_n]}} \bigg\} = d+1-\frac{dP-1}{d^2-1}.
\end{equation}
with equality for tight rank-1 IC-POVMs.

In contrast to projector-like observables, this lower bound is not bounded from above by a constant that is independent on the dimension $d$ and indeed one has that $\forall \rho$:
\begin{equation}
\min_{\bs \mu} \bigg\{ \overline{\overline{\mathrm{Var}[\hat{o}|\calO_n]}} \bigg\} 
\sim \calO (d), 
\qquad
d \rightarrow \infty.
\end{equation}

Comparing the average variances of~\cref{eq:average_variance_state_obs_witness} and~\cref{eq:average_variance_state_obs_pauli}, we can also write for general measurement frames $\bs \mu$
\begin{equation}
\overline{\overline{\mathrm{Var}[\hat{o}|\calO_n]}} = \frac{d^2}{d-1}  \overline{\overline{\mathrm{Var}[\hat{o}|\mathbb{P}_{\psi}]}}.
\end{equation}

\section{Optimal dual frame and rescaled frames}
\label{sec:rescaled_frames}

As discussed in~\cref{sec:shadow_tomography_formalism}, the non-rescaled canonical dual frame $\calF=\sum_b \mathbb{P}(\mu_b)$ is not, in general, the optimal choice of unbiased estimator.
Nonetheless, it can be interesting to notice that we can see the optimal dual frame as corresponding to the canonical dual frame computed with respect to a rescaled frame.
More precisely, the estimator $\tilde\mu^{(\rho)}$ introduced in~\cref{sec:estimators_derivation}, can be derived considering the rescaled frame with elements
\begin{equation}
    \mu_b^N \equiv \frac{\mu_b}{\sqrt{\langle\mu_b,\rho\rangle} }.
\end{equation}
The set of operators $\{\mu_b^N\}_b$ is a frame iff $\{\mu_b\}_b$ is also a frame, and the non-rescaled frame operator corresponding to $\{\mu_b^N\}_b$ is precisely the rescaled frame operator corresponding to $\{\mu_b\}_b$.

\parTitle{Frame operators for arbitrary rescalings}
We briefly show in this section the consequences of considering frames of operators defined in terms of rescaled POVM elements, for arbitrary rescalings. In particular, we show what the expansion of a generic state looks like using such formalism, and the associated unbiased estimators corresponding to each choice of rescaling.
These observations are not pivotal to the main results of the paper, but are presented here for the sake of completeness.

\parTitle{Definition of general rescaled measurement frames}
Consider a \textit{rescaled measurement frames} with elements $\mu_b/\sqrt{\alpha_b}$ for some set of positive real coefficients $\alpha_b$.
The associated non-rescaled frame operator is
\begin{equation}
    \calF_\alpha \equiv \sum_b \frac{\mathbb{P}(\mu_b)}{\alpha_b}.
\end{equation}
If $\mu_b^{(\alpha)\star}=\calF_\alpha^{-1}(\mu_b/\sqrt{\alpha_b})$ denotes the corresponding (non-rescaled) canonical dual frame, the associated decomposition of a state $\rho$ reads
\begin{equation}\label{eq:decomposition_rho_rescaled_frames}
    \rho = \sum_b \frac{\langle\mu_b,\rho\rangle}{\sqrt{\alpha_b} } \mu_b^{(\alpha)\star}
    = \sum_b \frac{\langle\mu_b,\rho\rangle}{\alpha_b} \calF_\alpha^{-1}(\mu_b).
\end{equation}
Recognising that $\langle \mu_b,\rho\rangle$ is a probability, we then define an unbiased estimator for the state as
\begin{equation}
    \hat f(b) \equiv \frac{1}{\sqrt{\alpha_b}} \mu_b^{(\alpha)\star},
\end{equation}
which thus satisfies $\mathbb{E}[\hat f]=\rho$.
Note that, in general, $\mu_b^{(\alpha)\star}\neq \sqrt{\alpha_b} \mu_b^\star$, and thus different frame scalings provide nontrivially different canonical estimators, albeit~\cref{eq:decomposition_rho_rescaled_frames} means that each set of operators $\{ \frac{1}{\sqrt{\alpha_b} }\mu_b^{(\alpha)\star} \}_b$ is a, generally non-canonical, valid dual frame of the non-rescaled measurement frame $\{\mu_b\}_b$.

\parTitle{Average error with rescaled frames}
The main usefulness of considering rescaled measurement frames is that the associated average $L_2$ error now reads
\begin{equation}
    \mathbb{E} \| \hat f-\rho\|_2^2
    = \mathbb{E} \trace(\hat f^2) - \trace(\rho^2),
\end{equation}
where
\begin{equation}
    \mathbb{E}\trace(\hat f^2) \equiv
    \sum_b \frac{\langle\mu_b,\rho\rangle}{\alpha_b} \trace((\mu_b^{(\alpha)\star})^2).
\end{equation}
Therefore, if we rescale the operators via $\alpha_b=\langle\mu_b,\rho\rangle$, we can write
\begin{equation}
    \mathbb{E} \trace(\hat f^2) = 
    \sum_b \trace((\mu_b^{(\alpha)\star})^2) =
    \trace(\calF_\alpha^{-1}).
\end{equation}
This simplifies the problem to searching for the measurement $\bs \mu$ that minimizes $\trace(\calF^{-1}_\alpha)$ using a given set of coefficients $\{\alpha_b\}_b$.

\section{Toy examples}
\label{app:toy_examples}

In this section we present a number of toy examples to better illustrate how the techniques set forth in the main text would be used in practice.

\subsection{Projective measurement}
Consider a simple single-qubit projective measurement: $\mu_0\equiv \mathbb{P}_0$ and $\mu\equiv \mathbb{P}_1$. The corresponding canonical frame superoperator, as per~\cref{eq:definition_canonical_estimator}, is
\begin{equation}
\begin{gathered}
    \calF_{I/2} = 2 [\mathbb{P}(\mathbb{P}_0) + \mathbb{P}(\mathbb{P}_1)] 
    = 2\begin{pmatrix}
        1 & 0 & 0 &0 \\
        0&0&0&0\\
        0&0&0&0\\
        0&0&0&1
    \end{pmatrix},
\end{gathered}
\end{equation}
where we represented the superoperator in the standard vectorized notation.
This $\calF_{I/2}$ is clearly singular, correspondingly to the POVM not being informationally complete. The associated estimator is not well-defined, correspondingly to the POVM not being a frame. Nonetheless, the general property $\calF_{I/2}(I/2)=I$, as per~\cref{eq:Frho_rho_equal_rho}, still holds, as directly verified observing that upon vectorization the identity operator $I$ becomes $\on{vec}(I)=(1,0,0,1)^T$. Similarly, the decomposition given in~\cref{eq:decomposition_FId_for_general_measurements} applies, and we can write
\begin{equation}\small
    \calF_{I/2} = 2 \mathbb{P}(I/\sqrt2) + \tilde\calF_{I/2},
    \quad
    \tilde\calF_{I/2} = \begin{pmatrix}
        1&0&0&-1 \\ 0&0&0&0 \\ 0&0&0&0 \\ -1&0&0&1
    \end{pmatrix},
\end{equation}
and we can directly verify that $\mathbb{P}(I/\sqrt2)$ and $\tilde\calF_{I/2}$ act on orthogonal spaces, and that $\tilde\calF_{I/2}$ acts on the space of traceless operators, as $\tilde\calF_{I/2}(Z)=2Z$, where $Z\equiv\mathbb{P}_0-\mathbb{P}_1$.

\subsection{Simple non-IC POVM}

Consider the following single-qubit POVM:
\begin{equation}
\begin{gathered}
    \mu_1 = \frac12\mathbb{P}_0, \quad \mu_2=\frac12\mathbb{P}_1, 
    \quad \mu_3 = \frac12\mathbb{P}_+, \quad \mu_4 = \frac12\mathbb{P}_-.
\end{gathered}
\end{equation}
Note that in vectorized notation we have $\mu_1=\frac12(1,0,0,0)$, $\mu_3=\frac14(1,1,1,1)^T$, etc.
The corresponding canonical frame superoperator is then
\begin{equation}
    \calF_{I/2} = 
    \frac12\begin{pmatrix}
        3 & 0 & 0 & 1\\
        0&1&1&0\\
        0&1&1&0\\
        1&0&0&3
    \end{pmatrix}.
\end{equation}
This has eigenvalues $\{2, 1, 1, 0\}$, and is therefore again singular, consistently with the POVM being again not informationally complete.
Note how the number of nonzero eigenvalues reflects the dimension of the span of the POVM, which is in this case larger than for the simple projective case.
The eigenvectors corresponding to the nonzero eigenvalues are $(1,0,0,1)^T$, $(1,0,0,-1)^T$, and $(0,1,1,0)^T$, respectively, which devectorizing correspond to the Pauli operators $I, Z$, and $X$. This is again consistent with the general statement that $\calF_{I/2}(I/2)=I$.
Note that in this example by defining the frame superoperator directly via~\cref{eq:frame_operator_def}, thus not introducing the rescaling factors used in~\cref{eq:definition_canonical_estimator}, the frame operator would have been $\calF_{I/2}/4$ instead.

\subsection{Example of IC-POVM}
As an example of a single-qubit IC-POVM, consider
\begin{equation}
\begin{gathered}
    \mu_1 = \frac13\mathbb{P}_0,
    \quad\mu_2 = \frac13\mathbb{P}_+,
    \quad\mu_3\equiv\frac13\mathbb{P}_R,\\
    \mu_4=I-\mu_1-\mu_2-\mu_3,
\end{gathered}
\end{equation}
with $\mathbb{P}_R = \ketbra{R}$ and $\ket{R} = (\ket{0}+i\ket{1})/\sqrt{2}$.

\parTitle{Frame operators and canonical estimator}
This POVM is informationally complete, and its corresponding frame operator is
\begin{equation}\small
    \calF_{I/2} = \frac1{18}\begin{pmatrix}
        22 & 1+i & 1-i & 14\\
        1-i & 8 & -2i & -1+i \\
        1+i & 2i & 8 & -1-i \\
        14 & -1-i & -1+i & 22
    \end{pmatrix},
\end{equation}
whose eigenvalues are
$2, 2/3,1/3,1/3$.
The actual matrix representation of the frame operator depends on the choice of operator basis. The above representation corresponds to a standard choice of operatorial basis with elements $\{\ketbra{i}{j}\}_{i,j}$.
Another possibility is to represent the operator in a basis of Hermitian operators, such as $\{I/\sqrt2,X/\sqrt2,Y/\sqrt2,Z/\sqrt2\}$. With this choice, we get instead
\begin{equation}
    \calF_{I/2} =
    \begin{pmatrix}
        2 & 0 & 0 & 0 \\
        0&4/9&1/9&1/9\\
        0&1/9&4/9&1/9\\
        0&1/9&1/9&4/9
    \end{pmatrix},
\end{equation}
which makes some of underlying structure more transparent.
As always, the first eigenvalue corresponds to the $I$ eigenvector, that is, the general property $\calF_{I/2}(I)=2I$.
The remaining eigenvalues are eigenvalues of $\tilde\calF_{I/2}$.
In particular, the eigenvectors corresponding to the eigenvalues $2/3,1/3,1/3$, and thus also the eigenvectors of $\tilde\calF_{I/2}$, are the operators $X+Y+Z$, $X-Z$, and $X+Z-2Y$, respectively.
We can now compute the canonical estimator elements $\tilde\mu_b^{\rm can}$, which work out to be
\begin{equation}\small
\begin{gathered}
    \tilde\mu_1^{\rm can} = \frac12(I-X-Y+5Z),
    \quad
    \tilde\mu_2^{\rm can} = \frac12(I + 5X-Y - Z), \\
    \tilde\mu_3^{\rm can} = \frac12(I - X+ 5 Y - Z), \quad
    \tilde\mu_4^{\rm can} = \frac12(I - X - Y - Z).
\end{gathered}
\end{equation}
These then provide unbiased estimators to estimate arbitrary observables.
For example, if the target observable is the Pauli matrix, $\calO=Z$, then the observable estimator would be $\hat o$ such that
$\hat o(b)=\langle \calO,\tilde\mu_b^{\rm can}\rangle$, whose values are
\begin{equation}
    \hat o(1) = 5,
    \qquad
    \hat o(2) = \hat o(3) = \hat o(4) = -1.
\end{equation}
Being the POVM minimal, meaning the number of outcomes equals $d^2$, the number required to have informational completeness, the POVM elements are also in this case linearly independent. This implies that there is a single possible choice of dual frame, and therefore a single choice of estimator.
In other words, performing similar calculations using the non-rescaled frame operators, will produce the same exact estimators in this case.

\parTitle{Assessment of estimator variances}
We can then use~\cref{eq:variance_observable_estimator} to compute the variances in different scenarios. For example, if $\rho=\mathbb{P}_0$ and $\calO=Z$, then
\begin{equation}
\begin{gathered}
    \on{Var}[\hat o|\PP_0,Z] =
    \langle \calO,\calC_{\PP_0}(\calO)\rangle \\
    = \left[5^2\frac{1}{3} + (-1)^2 \left(1-\frac1{3} \right)\right] - 1 = 8,
\end{gathered}
\end{equation}
where $\calC_{\mathbb{P}_0}$ is the MSE matrix, as defined in~\cref{eq:mse_matrix_definition}, computed using the canonical estimator $\tilde\mu_b^{\rm can}$.
If, on the other hand, we have $\rho=\mathbb{P}_1$, then
\begin{equation}
    \on{Var}[\hat o|\mathbb{P}_1,X]
    = \on{Var}[\hat o|\mathbb{P}_1,Y] = 5,
\end{equation}
but $\on{Var}[\hat o|\mathbb{P}_1,Z]=0$, consistently with the first outcome being the only one that gives $\hat o(1)=5$, and this outcome having zero probability due to $\langle \mu_1,\mathbb{P}_1\rangle=0$.
We can also gain a more general understanding of how the variance changed with the input state using the $A$ operator defined in~\cref{eq:definition_A_op}.
For example, if $\calO=X$, this equals
\begin{equation}
    A = \sum_b \langle X,\tilde\mu_b^{\rm can}\rangle \mu_b =
    \begin{pmatrix}
        5&4\\4&5
    \end{pmatrix}.
\end{equation}
This operator has eigenvalues $9,1$, which immediately tells us that $1\le \mathbb{E}[\hat o^2] \le 9$, and thus $0\le\on{Var}[\hat o|\rho,X] \le 9$.
In particular, the eigenvector of $A$ corresponding to the eigenvalue $+1$ is $(1,-1)^T$, which tells us that the state $\rho=\mathbb{P}_-$ is such that $\langle A,\mathbb{P}_-\rangle=1$, and because $\langle X,\mathbb{P}_-\rangle^2=1$, we conclude that $\on{Var}[\hat o|\mathbb{P}_-,X]=0$.

\parTitle{Bounds on the average variance}
To work with averaged variance, we can use~\cref{eq:averaged_variance_general_formula,eq:averaged_variance_vs_eigenvalues,eq:averaged_variance_final_expr}, which immediately tell us that the possible values of the averaged variance depend on the eigenvalues of $\tilde\calF_{I/2}$. As shown above, in the case we are studying, these eigenvalues are $3,3,3/2$. Sticking to pure states for simplicity, we thus get the general bounds for the averaged variance in this example as:
\begin{equation}\label{eq:bound_avgvar_thirdexapmle}
    2.67 \simeq \frac83 \le \frac{\overline{\on{Var}[\hat o|\calO]}}{V} \le \frac{17}{3} \simeq 5.67.
\end{equation}

\parTitle{Consistency with general bounds on average variance}
Finally, we can also attempt to directly verify the consistency of the general bounds provided in~\cref{eq:worst-case-variance-bound}.
Working out explicitly the various terms for our canonical frame operator we find $a=10/3$, $b=14/3$, and
\begin{equation}
    \lambda_1^* =
    \frac{10-\sqrt{13}}{9}
    \simeq 0.71,
\end{equation}
and thus the bound reads, considering pure states for simplicity,
\begin{equation}
    \max_{\calO} \frac{\overline{\on{Var}[\hat o|\calO]}}{V} \ge
    \frac{2}{\lambda_1^*} - \frac13
    \simeq 2.48.
\end{equation}
This is consistent with~\cref{eq:bound_avgvar_thirdexapmle}, because
$2.48< 17/3$.
This tells us that there are better choices of measurement which produce frame superoperators compatible with the given values of $a,b$, that give much better worst-case average variance.

\bibliography{main}



\end{document}